\documentstyle[epsf,preprint,eqsecnum,,aps]{revtex}
\begin{document}
\draft
\title{Density Fluctuation Mediated Superconductivity}

\author{P. Monthoux and G.G. Lonzarich}
\address{Cavendish Laboratory, University of Cambridge
\\Madingley Road, Cambridge CB3 0HE, United Kingdom}
\date{\today}
\maketitle

\begin{abstract}

We compare predictions of the mean-field theory of superconductivity for 
metallic systems on the border of a density instability for cubic and 
tetragonal lattices.  The calculations are based on the parameterisation 
of an effective interaction arising from the exchange of density
fluctuations and assume that a single band is relevant for 
superconductivity. The results show that for comparable model parameters, 
density fluctuation mediated pairing is more robust in quasi-two dimensions 
than in three dimensions, and that the robustness of pairing increases 
gradually as one goes from a cubic structure to a more and more anisotropic 
tetragonal structure. We also find that the robustness of density
fluctuation mediated pairing can depend sensitively on the incipient
ordering wavevector.  We discuss the similarities and differences
bewteen the mean-field theories of superconductivity for density mediated
and magnetically mediated pairing.

\end{abstract}


\narrowtext

\section{Introduction}

Soon after the development of the BCS\cite{BCS} theory of superconductivity 
it was realized that Cooper instabilities could arise not only from the 
exchange of phonons, but also from overscreening of the Coulomb interaction 
in a model Fermi liquid without ion dynamics.  Due to the sharpness of the 
Fermi surface, the generalized spin and charge susceptibilities exhibit 
Friedel oscillations in space.  Kohn and Luttinger\cite{KL} argued that 
oscillations of a similar origin can show up in the effective interaction 
between quasiparticles.  Keeping all diagrams up to second order in a 
model bare fermion-fermion interaction, they also found an induced attraction 
which does not rely on the presence of the Friedel oscillations.  They
demonstrated that it was possible in principle to construct an anisotropic
Cooper state that sampled mainly the attractive regions of the effective
pairing interaction.

The Kohn-Luttinger model applies where the susceptibilities or response
functions are adequately represented by the bare Lindhard function for a
homogeneous electron gas for which the oscillations are weak and hence the
calculated superconducting transition temperatures $T_c$ turn out to be
typically well below the experimentally accessible range.  As a charge or
spin instability is approached, however, we expect the corresponding
response function and its real space attractive regions to become enhanced.
Provided that it is possible to match the Cooper pair state to the
attractive regions, $T_c$ may rise to the experimentally accessible range.  An
extension of the Kohn-Luttinger theory would naturally lead to spin
dependent quasiparticle interactions.  Current models of magnetically
mediated superconductivity focus on such magnetic interactions which are
expected to dominate on the border of magnetic long-range order.

It has been shown that this magnetic interaction treated at the mean-field 
level can produce anomalous normal state properties and superconducting 
instabilities to anisotropic pairing states.  It correctly predicted the 
symmetry of the Cooper state in the copper oxide 
superconductors\cite{DwavePrediction} and is consistent with 
spin-triplet p-wave pairing in superfluid $^3He$ [for a 
recent review see, e.g., ref.~\cite{Dobbs}].  One also gets the correct 
order of magnitude of the superconducting and superfluid transition 
temperature $T_c$ when the model parameters are inferred from experiments 
in the normal state of the above systems.  There is growing evidence 
that the magnetic interaction model may be relevant to other materials 
on the border of magnetism.

In our previous work\cite{ML1,ML2,ML3} we focussed on clarifying the 
general features of the magnetic interaction model.  The latter may be 
relevant to understanding the superconductivity recently discovered, 
for example, on the border of antiferromagnetism in systems such as cubic 
$CeIn_3$\cite{Walker} and its tetragonal counterpart $CeRhIn_5$\cite{Hegger} 
and on the border of ferromagnetism in $UGe_2$\cite{Saxena}, $URhGe$\cite{Huxley} 
and $ZrZn_2$\cite{Pfleiderer}.

In contrast to the conventional phonon-mediated interaction, which is
usually taken to be local in space but non-local in time, the magnetic
interaction is non-local in both space and time.  The non-locality in space
leads to anisotropic pairing states whose nature can be acutely sensitive to
details of the lattice and electronic structure, and the form of the
quasiparticle interaction .  For simplicity, in references\cite{ML1,ML2,ML3} 
we considered only a simple cubic or tetragonal crystal structure, a 
single tight-binding energy band and a magnetic interaction treated at 
the mean-field level. 

One crucial aspect of the magnetic interaction is the vector nature of the
spin degree of freedom.  At first sight it might appear that the
longitudinal and transverse phonons that mediate the usual lattice
interaction would be analogous to the longitudinal and transverse spin
fluctuations that mediate the magnetic interaction.  However, the latter
interaction depends on the relative spin orientation of the interacting 
particles and hence has a different sign and magnitude for the spin-singlet 
and spin-triplet Cooper states.  By contrast, the conventional phonon 
mediated interaction is oblivious to the spin degree of freedom of 
the quasiparticles.

One of the consequences is that on the border of ferromagnetism, the
magnetic interaction is typically only attractive in the spin-triplet
channel.  In that case, only the longitudinal spin fluctuations contributing
to pairing while all three contribute to the self-interaction that tends to
be pair breaking.  This disadvantage can be mitigated in systems with strong
magnetic anisotropy in that the effect of the transverse spin fluctuations
on the self-interaction would be suppressed while the strength of the
pairing interaction arising from longitudinal spin fluctuations need not be
reduced.  By contrast to isotropic magnetic pairing in the spin-triplet
channel, in conventional phonon mediated superconductivity and in magnetic
pairing in the spin-singlet channel, all modes contribute both to
the pairing and to the self-interaction effects.

We have also found that for the model considered in references\cite{ML1,ML2,ML3} 
the robustness of magnetic pairing increases gradually as one goes from a cubic
to a more and more anisotropic tetragonal structure under otherwise similar
conditions.  This is due to the increase with growing anisotropy of the
density of states of both the quasiparticles and of the magnetic
fluctuations that mediate the quasiparticle interaction.  One expects and
calculations presented in this paper show that this result carries over to
other pairing mechanisms treated at the same level of approximation as in
references\cite{ML1,ML2,ML3}.

To further our understanding of the conditions favorable to robust pairing
it would seem natural to carry out similar types of analyses of
superconductivity on the border of other types of instabilities.  We
consider the possibility of pairing near instabilities signalled by the
divergence of a particle density response function.  This could include in
principle structural instabilities characterized by the softening of phonons
in some regions of the Brillouin zone.  The induced interaction produced by
these soft phonons, in contrast to conventional phonons, is non-local in
space.  Therefore, one could expect some similarities to the magnetic
pairing problem studied in references\cite{ML1,ML2,ML3}.

A density response function may also be expected to be strongly enhanced on
the border of a charge density wave (CDW) transition, a stripe instability
and an $\alpha-\gamma$ or valence instability.  The appropriate density response
function may be expected to become large at a wavevector near the Brillouin
zone boundary for a CDW, at small but finite wavevectors for stripes and at
zero wavevector near the $\alpha-\gamma$ transition (at which the 
structure of the unit cell remains the same, but its volume changes).

We note that if the density transition happens to be strongly first order,
the appropriate density response function and hence the associated
quasiparticle interaction may not be sufficiently enhanced to lead to an
observable superconducting phase.  This is particularly relevant to 
the $\alpha-\gamma$ transition commonly found in heavy fermion systems, 
which is similar to the liquid-gas transition in that it is first order 
except at the critical end point.  When the latter is at a temperature well 
above the expected temperature scale for pairing, the enhanced density 
fluctuations associated with the $\alpha-\gamma$ transition are unlikely 
to produce superconductivity.  In the temperature region near the critical 
end point when density fluctuations are strong, superconductivity would be 
suppressed by thermal fluctuations, while in the low temperature regime the 
density fluctuations are too weak because of the strong first order character 
of the density transition.  This could explain the absence of superconductivity 
in $CeNi$ where the critical end point is around room temperature, but the 
existence of superconductivity in $CeCu_2Si_2$ and $CeCu_2Ge_2$ at high 
pressures where a corresponding critical end point is believed to exist at 
low temperatures or may be just suppressed.

\section{Model}

We consider quasiparticles in a simple tetragonal lattice described by a 
dispersion relation

\begin{eqnarray}
\epsilon_{\bf p} & = & -2t(\cos(p_x) + \cos(p_y) + \alpha_t\cos(p_z)) 
\nonumber \\ & - & 4t'(\cos(p_x)\cos(p_y) + \alpha_t\cos(p_x)\cos(p_z) 
+ \alpha_t\cos(p_y)\cos(p_z))
\label{eps}
\end{eqnarray}

\noindent with hopping matrix elements $t$ and $t'$. $\alpha_t$ represents 
the electronic structure anisotropy along the z direction.  
$\alpha_t = 0$  corresponds to the 2D square lattice while 
$\alpha_t = 1$ corresponds to the 3D cubic lattice.  For 
simplicity, we measure all lengths in units of the respective lattice 
spacing.  In order to reduce the number of independent parameters, we take 
$t' = 0.45 t$ and a band filling factor $n = 1.1$ as in our earlier 
work\cite{ML1,ML2,ML3}.

The effective interaction between quasiparticles is taken to be the induced
density-density interaction and is defined in terms of a coupling constant g
and a generalized density susceptibility, which is assumed to have a simple
analytical form consistent with the symmetry of the lattice,

\begin{equation}
\chi({\bf q},\omega) = {1\over N_{\bf q_0}}\sum_{\bf q_0}
{\chi_0\kappa_0^2\over \kappa^2 + \Delta({\bf q})
- i{\omega\over \eta(\widehat{q})}}
\label{chiML}
\end{equation}

\noindent where $\kappa$ and $\kappa_0$ are the correlation wavevectors or 
inverse correlation lengths in units of the lattice spacing in the basal 
plane, with and without strong density correlations, respectively. 
The function $\Delta({\bf q})$, in Eq.~(\ref{chiML}), is defined as

\begin{equation}
\Delta({\bf q}) = (4 + 2\alpha_d) - 
2(\cos(q_x-q_{0x})+\cos(q_y-q_{0y})+\alpha_d\cos(q_z-q_{0z})) 
\label{dqdef}
\end{equation}

\noindent where $\alpha_d$ parameterises the density anisotropy. $\alpha_d = 0$ 
corresponds to quasi-2D density correlations and $\alpha_d = 1$ 
corresponds to 3D density correlations. The sum in Eq.~(\ref{chiML}) is over
all the symmetry related vectors ${\bf q_0}$, with $N_{\bf q_0}$ the number
of such vectors. In the following, we only explicitly write one of the vectors. 
It should be understood that when we say that the incipient wavevector is, 
for example, ${\bf q_0} = [\pi/4,0]$, it is implied that the density response 
function peaks at the four wavevectors $[\pm \pi/4,0],[0,\pm\pi/4]$. 
The parameter $\eta(\widehat{q})$ in Eq.~(\ref{chiML}) is defined as

\begin{eqnarray}
\eta(\widehat{q}) & = & T_{DF}\widehat{q}^n
\label{eta} \\
\widehat{q}^2 & = & (4 + 2\alpha_d) - 
2(\cos(q_x)+\cos(q_y)+\alpha_d\cos(q_z)) 
\label{qdef}
\end{eqnarray}

\noindent where $T_{DF}$ is a characteristic density fluctuation temperature.
In Eq.~(\ref{eta}), the exponent $n=1$ if the density fluctuations are such 
that the total density is conserved and $n=0$ otherwise. We note that the pole 
of the density response function Eq.~(\ref{chiML}) is purely imaginary and 
therefore the density fluctuations we consider are overdamped.  This is 
believed to apply on the border of CDW, stripe and valence instabilities, 
but not typically for lattice density fluctuations for which the poles of 
the density response must have a non-negligible real component.  The latter 
would require the inclusion of an $\omega^2$ term with real coefficient 
in the denominator of Eq.~(\ref{chiML}).

In addition to the induced density interaction, we include an on-site
Coulomb repulsion $I$.  In the large $I$ limit, the Cooper pair state 
vanishes when the interacting quasiparticles are on the same site and 
thus conventional isotropic s-wave pairing is excluded.

We note that in the corresponding problem of magnetic pairing the effective
interaction is repulsive when the two interacting quasiparticles are on the
same site in the spin-singlet channel.  It is, however, attractive in the
spin-triplet channel, but this is irrelevant since the required spatial
antisymmetry of the pair state means that the two quasiparticles have zero
probability of occupying the same site simultaneously.

A complete description of the model, the Eliashberg equations for the 
superconducting transition temperature and their method of solution 
can be found in the appendix.

\section{Comparison of the Density and Magnetic Pairing Interactions}

Our assumed form of the density response function is similar to that of the
generalized magnetic susceptibility used in our previous papers.  However,
there is a crucial difference in that the effective magnetic interaction
depends on the relative orientation of the spins of the two quasiparticles
through the factor ${\bf \sigma_1\cdot\sigma_2}$.  In the spin-singlet state 
the expectation value of ${\bf \sigma_1\cdot\sigma_2}$ gives a factor of -3.  
When the interaction is oscillatory in real space, this sign change leads 
to an interchange of attractive and repulsive regions.  Since one must choose
a pair state in which the quasiparticles mainly sample the attractive region 
of the interaction, the sign inversion implies a change in the symmetry of 
the Cooper pair state as illustrated in Figs. 1a and 2a for the cases of incipient 
ordering wavevectors ${\bf q_0} =[\pi,\pi]$ and $[\pi,0]$ in a square lattice.  
For the case of small ${\bf q_0}$, where the oscillations are essentially 
irrelevant in our model, the density interaction is attractive in real space 
for both the spin-singlet and spin-triplet states, but the magnetic interaction 
is attractive solely for the spin-triplet state for which the expectation value 
of ${\bf \sigma_1\cdot\sigma_2}$ is +1.

In our model for the generalized magnetic susceptibility we have assumed
that the overall magnetization is conserved and hence $\eta({\widehat q})$ vanishes 
as ${\bf q} \rightarrow 0$. This leads to greater incoherent scattering for 
a nearly ferromagnetic than antiferromagnetic metal, and hence to a reduction of 
$T_c$ on the border of ferromagnetism.  If the fluctuations of the density 
are quasi-local as in some models of valence fluctuations\cite{Miyake}, 
then $\eta({\widehat q})$ does not vanish at small ${\bf q}$.  This corresponds 
to the case $n = 0$ in Eq.~(\ref{eta}).  If ${\bf q_0}$ is sufficiently
far away from the origin in the Brillouin zone, the precise value of $n$ is not
expected to affect the calculated $T_c$.  Since we are not going to consider
the limit ${\bf q_0} = 0$, for simplicity we take $n = 1$ as in the case of the
magnetic interaction.

\section{Results}

\subsection{Quasi 2D: ${\bf q_0} = [\pi,\pi]$ and $[\pi/m,0]$ with $m = 1, 2, 4$}

The dimensionless parameters at our disposal are $g^2\chi_0/t$, $T_{DF}/t$, 
$\kappa_0$ and $\kappa$. For comparison with results of our earlier work 
for the case of the magnetic interaction, we take $T_{DF} = 2t/3$ and 
$\kappa_0^2 = 12$.  In 2D, this $T_{DF}$ corresponds to about 1000 K for a 
bandwidth of 1 eV, while our choice of $\kappa_0^2$ is a representative value.  
We note that $\kappa_0^2/\kappa^2$ represents the density susceptibility 
enhancement factor, analogous to the Stoner factor in the case of the 
magnetic interaction.

The results of our numerical calculations of the mean-field critical
temperature $T_c$ as a function of $g^2\chi_0/t$ and of $\kappa^2$ is shown 
in Fig. 1 for ${\bf q_0} = [\pi,\pi]$ in which the Cooper pair state has 
$d_{xy}$ symmetry.  The nodal lines of this state in real space are illustrated 
in Fig. 1a, which also depicts the static density interaction seen by one of 
the quasiparticles given that the other is at the origin.  For values of the 
dimensionless coupling parameter $g^2\chi_0/t$ corresponding to the Random Phase 
Approximation (of order 10), $T_c$ is found to drop very rapidly as one goes 
away from the instability, i.e with increasing $\kappa^2$.

The corresponding plots for the cases ${\bf q_0} = [\pi/m,0]$, 
where $m$ = 1, 2 and 4 are shown in Fig. 2, 3 and 4.  In contrast to the 
case ${\bf q_0} = [\pi,\pi]$, the next-nearest-neighbor interaction for 
${\bf q_0} = [\pi,0]$ is repulsive.  This requires nodal lines along 
the diagonal, and  hence the $d_{x^2-y^2}$ instead of $d_{xy}$ symmetry.  
As shown in Fig. 2a, the nearest-neighbor interaction vanishes for 
the special case ${\bf q_0} = [\pi,0]$ and the leading attraction comes 
from third-nearest-neighbors.  This explains why pairing is not as 
robust in this case compared with the case ${\bf q_0} = [\pi,\pi]$.

As seen from Figs 3a and 4a, the strength of the nearest-neighbor
attraction increases as ${\bf q_0}$ gets smaller, which correlates 
with the increased robustness of $T_c$.

As ${\bf q_0}$ decreases the density interaction can also be attractive 
for other pairing states.  In order to avoid the on-site Coulomb interaction, 
one could use the $d_{xy}$ state since the next-nearest-neighbor interaction 
is attractive for sufficiently small ${\bf q_0}$.  But since the 
$d_{x^2-y^2}$ state picks the nearest-neighbor attraction, which is 
dominant, it is expected to be the favored state. For small ${\bf q_0}$,
the density interaction is also attractive in the spin-triplet channel
for a $p_x$ or $p_y$ Cooper state. This state picks two out of the four
nearest-neighbor attractive sites, instead of all four for the $d_{x^2-y^2}$ 
state. However, the $p_x$ or $p_y$ state also picks the attraction on all 
four next-nearest-neighbor sites where the $d_{x^2-y^2}$ state vanishes.
It is thus not immediately obvious in that case which of the two pairing
states has the highest $T_c$. Fig. 5 shows the Eliahsberg superconducting 
transition temperature one obtains for the spin-triplet $p_x$ and spin-singlet
$d_{x^2-y^2}$ states as a function of the correlation wavevector $\kappa^2$
for $g^2\chi_0/t = 10$. The plot shows that the $d_{x^2-y^2}$ is the favored 
case and we have found this to be true for the range of values of $\kappa^2$
and $g^2\chi_0/t$ studied in this paper.

\subsection{3D: ${\bf q_0} = [\pi,\pi,\pi]$, ${\bf q_0} = [\pi/4,0,0]$}

The results of the numerical calculations in 3D are shown in Figs 6 and 7
for ${\bf q_0} = [\pi,\pi,\pi]$ in the $d_{xy}$ Cooper state and 
${\bf q_0} = [\pi/4,0,0]$ in the $d_{x^2-y^2}$ Cooper state, respectively. 
The pairing for ${\bf q_0} = [\pi/4,0,0]$ is more robust than for 
${\bf q_0} = [\pi,\pi,\pi]$ since the dominant attraction comes from 
nearest neighbor in the former case rather than next-nearest-neighbor 
as in the latter case. For ${\bf q_0} = [\pi,\pi,\pi]$, pairing 
is less robust in 3D than for the corresponding quasi-2D case shown 
in Fig. 1 for all coupling constants. In the case ${\bf q_0} = [\pi/4,0,0]$, 
pairing is more robust in quasi-2D (Fig. 4) than in the corresponding 3D case 
for weak to intermediate coupling. At strong coupling, however,
pairing is more robust in 3D, but coupling constants $g^2\chi_0/t$ 
in the 20-60 range are less physically realistic.

\subsection{Crossover from 3D to quasi-2D:  Tetragonal lattice with 
${\bf q_0} = [\pi,\pi,\pi]$, ${\bf q_0} = [\pi/4,0,0]$}

The calculated $T_c$ as a function of the electronic and density response
anisotropy parameters $\alpha_t$ and $\alpha_d$, respectively, are shown in 
Figs. 8 and 9 for representative values of the parameters $\kappa^2$ and 
$g^2\chi_0/t$.  The results reported in sections A and B correspond to 
the quasi-2D case $\alpha_t = \alpha_d = 0$ and to the
3D case $\alpha_t = \alpha_d = 1$.

For ${\bf q_0} = [\pi,\pi,\pi]$, shown in Fig. 8, we find that $T_c$ 
increases gradually and monotonically as the system becomes more and 
more anisotropic in the density interaction.  We also note that the 
effect of the electronic anisotropy is much less pronounced. In the
case of an incipient ordering wavevector ${\bf q_0} = [\pi/4,0,0]$,
Fig. 9 shows that $T_c$ is maximum for an anisotropy parameter
$\alpha_d$ between 0 and 1, namely for an anisotropic albeit not 
quasi-2D in the density interaction. Also note that in the 
${\bf q_0} = [\pi/4,0,0]$ case, $T_c$ depends more strongly on 
the electronic anisotropy parameter $\alpha_t$ than for 
${\bf q_0} = [\pi,\pi,\pi]$.

\section{Discussion}

\subsection{Role of Real Space Oscillations in the Quasiparticle Interaction}

When the wavevector ${\bf q_0}$ at which the density response function 
is a maximum lies near the Brillouin zone boundary the quasiparticle 
interaction has short-range real-space oscillations.  As a consequence, 
the robustness of the pairing depends sensitively on whether one can 
construct a Cooper pair state from quasiparticle states near the Fermi 
surface such that given one quasiparticle is located at the origin, 
the probability of finding the second one in regions where the interaction 
is repulsive is minimized.  For the case ${\bf q_0} = [\pi,\pi]$, this 
forces us to consider a Cooper state with nodes along the principal 
(x and y) axes (see Fig. 1a).

In the density interaction channel, the dominant attraction comes from the
next-nearest-neighbor sites and is typically much weaker than the dominant
nearest-neighbor attraction for spin-singlet magnetic pairing for the same
wavevector ${\bf q_0} = [\pi,\pi]$.  This explains why for this wavevector 
${\bf q_0}$, pairing is not as robust for the density interaction as for 
the magnetic interaction under otherwise similar conditions.

One might think that if there were a wavevector ${\bf q_0}$ such that the 
interaction is attractive at the nearest-neighbor sites, one could 
achieve pairing in the density channel to the same degree of robustness 
as in the spin-singlet magnetic channel for ${\bf q_0} = [\pi,\pi]$.  
A potential candidate wavevector is ${\bf q_0} = [\pi,0]$ since by 
rotating the wavevector one would rotate the oscillation pattern in 
real space.  However, the oscillations one obtains via Eq.~(\ref{chiML})
are superpositions of oscillations running along the x and y 
directions coming from the symmetry related components with wavevectors 
$[\pi,0]$ and $[0,\pi]$.  These oscillations perfectly cancel at the odd sites 
(see Fig. 2a), and in particular at nearest-neighbor sites.  The dominant 
attraction arises from the third-nearest-neighbors, and thus contrary 
to naive expectations the case with ${\bf q_0} = [\pi,0]$ leads to even 
weaker pairing than with ${\bf q_0} = [\pi,\pi]$. Note that for the 
corresponding spin-singlet magnetic pairing for ${\bf q_0} = [\pi,0]$,
because of the inversion of the sign of the interaction due to the
spin factor ${\bf \sigma_1\cdot\sigma_2}$, the dominant attraction would now 
come from the next-nearest-neighbor sites.  In this case, pairing would 
be more robust in the magnetic than in the density channel for 
${\bf q_0} = [\pi,0]$, but still not as favorable as the magnetic spin 
singlet channel for ${\bf q_0} = [\pi,\pi]$.

The robustness of density pairing for the simple tetragonal lattice is
optimized for ${\bf q_0}$ close to the Brillouin zone center since in 
that case the interaction at all neighboring sites is maximally attractive 
(see Fig. 4a).  In order to avoid the on-site Coulomb repulsion, the pairing 
state which is the solution of the gap equation (Appendix) vanishes at 
the origin and its symmetry is of the form $p_x$, $p_y$, $d_{xy}$ or 
$d_{x^2-y^2}$ .  Since the  $d_{x^2-y^2}$ state has maximum
amplitude at nearest-neighbor sites, it has the highest $T_c$.

By contrast to the case of the magnetic interaction where the most robust
pairing was shown in our previous work\cite{ML1} to arise for 
${\bf q_0} = [\pi,\pi]$, in the density channel our results indicate 
that the optimal case is for ${\bf q_0}$ near the Brillouin zone center.  
Since the symmetry of the Cooper state is the same in both cases, this 
would suggest that still stronger pairing should arise when the system 
is on the border of both a magnetic instability with ${\bf q_0}$ near 
$[\pi,\pi]$ and a density instability with low ${\bf q_0}$.  In that 
case, the two pairing mechanisms would reinforce each other rather 
than compete.

This observation may be very relevant to the superconductivity in 
f-electron compounds such as $CeCu_2Si_2$ and $CeCu_2Ge_2$.  In these 
systems the superconductivity extends over a region in pressure containing 
both an antiferromagnetic and a valence instability.  What is special 
about these materials is that the critical end point of the latter
instability lies at unusually low temperatures or is incipient, which 
means that density fluctuations are expected to be important in 
the temperature regime where superconductivity is observed.  
Since the two instabilities do not occur at the same pressure one would 
expect that near the magnetic instability the pairing would be 
dominated by the magnetic channel and as the pressure is increased 
that it would cross over to a regime dominated by the density
channel.

When the two instabilities are sufficiently widely separated, one might
expect to see two distinct superconducting domes, one centred near the
magnetic instability and the other near the density instability.  A double
domed superconducting temperature-pressure phase diagram has in fact been
observed in $CeCu_2Si_2$ and $CeCu_2Ge_2$ systems\cite{Jaccard,Yuan} 
and in $CeNi_2Ge_2$\cite{Grosche}. Some of these experimental findings 
have been interpreted in terms of the effects of magnetic and valence 
fluctuations\cite{Miyake}.

The overall scale of $T_c$ is set by the characteristic temperature of 
magnetic and density fluctuations which tends to be below 100 K in the 
above f-systems.  One way to increase the value of $T_c$ is to increase 
these characteristic temperatures.  This could be achieved by looking 
for analogous d-metal systems with broader electron bands.  The
antiferromagnetic and stripe fluctuations in the cuprates may be an 
example where magnetic\cite{DwavePrediction} and density 
fluctuations\cite{DiCastro} with high characteristic 
temperature scales reinforce to produce high temperature superconductivity.

\subsection{Role of Crystalline Anisotropy}

The numerical results show that the robustness of density mediated
superconductivity increases gradually and monotonically as one goes 
from a cubic to a more anisotropic tetragonal structure for 
${\bf q_0} = [\pi,\pi,\pi]$ and that $T_c$ is optimum for an 
anisotropic albeit not quasi-2D system for ${\bf q_0} = [\pi/4,0,0]$.  
One can partly understand this result by looking at the evolution of 
the density interaction in real space with increasing anisotropy as 
illustrated qualitatively in Figs. 10 and 11 for ${\bf q_0} = [\pi,\pi,\pi]$ 
and ${\bf q_0} = [\pi/4,0,0]$, respectively.  
We see that the attraction in the basal plane gets enhanced as one goes 
from the cubic to a more anisotropic tetragonal lattice.  This enhancement 
is the consequence of the increase of the phase space of soft density 
fluctuations as one goes from a cubic to a quasi-2D structure. 
Note that for ${\bf q_0} = [\pi/4,0,0]$, the model pairing potential
is not continuous at $\alpha_d=0$ since the number of peaks of the
density response goes from four (at $[\pm\pi/4,0],[0,\pm\pi/4]$) in 
stricly 2D to six (at $[\pm\pi/4,0,0],[0,\pm\pi/4,0],
[0,0,\pm\pi/4]$) for $\alpha_d>0$. Other than that, our 
model potential varies smoothly with the tetragonal distortion, 
parameterised by $\alpha_d$ in Figs. 10 and 11, and it is clear that 
this effect grows gradually with increasing separation between 
the basal planes.  In our Eliashberg calculations, mass renormalization
effects, which tend to suppress $T_c$, also increase as one goes to
a more and more anisotropic crystal structure. Our results thus depend
on the interplay between the strengths of the pairing interaction
and mass renormalization, and the fact that the maximum 
$T_c$ in the case ${\bf q_0} = [\pi/4,0,0]$ occurs for anisotropic 
but not quasi-2D systems reflects the delicate balance between these
opposing effects.
The above given phase space argument is similar to that used to 
explain the increased robustness of magnetic pairing with increasing
lattice anisotropy and, hence, as anticipated in Ref.\cite{ML3}, 
carries over to other pairing mechanisms treated at the one-loop 
mean-field level.  Another potential benefit of going to a more 
anisotropic crystal structure is the narrowing of the electronic 
band and the associated increase in the electronic density of states.  
Our results show that in the case of ${\bf q_0} = [\pi,\pi,\pi]$ and the 
model parameters considered, this does not play the dominant role.
However, for an incipient ordering wavevector ${\bf q_0} = [\pi/4,0,0]$,
the increase in the electronic density of states with increased lattice 
anisotropy plays a more important role. This effect could also be 
sensitive to details of the electronic and crystal structure not 
considered here. 

The calculations presented in this paper and in our previous 
work\cite{ML2,ML3} show that, in the majority of cases considered,
the lattice anisotropy increases the robustness of magnetic and density
pairing in the mean-field approximation.  Superconducting phase 
fluctuations which are not included in this approximation may 
be expected to suppress $T_c$ in the 2D limit.  Therefore, 
in practice, one would think that the most favorable case for 
magnetic or density pairing is that of strong, but not extreme, 
anisotropy where the effect of the superconducing phase fluctuations 
are typically weak. This is to be contrasted with the effect 
of order parameter fluctuations on magnetic and density transitions 
that can be large in metals even in 3D and more so in 2D.  In the 
case of the density transition, even a small lattice anisotropy 
and the resulting increase in the order parameter fluctuations 
can lower the critical end point significantly.  By weakening the 
first order transition at low temperatures, this would enhance 
the density fluctuations that mediate the pairing on the border 
of the density instability and lead to a superconducting phase.  

The importance of crystalline anisotropy in enhancing the superconducting 
$T_c$ on the border of antiferromagnetism has been dramatically 
demonstrated in going from the simple cubic system $CeIn_3$\cite{Walker} 
to the related tetragonal compounds $CeMIn_5$\cite{Hegger} where $M = Co$, 
$Rh$ and $Ir$, as correctly anticipated by our earlier model 
calculations of magnetic pairing\cite{ML1,ML2,ML3}.  

In addition to an antiferromagnetic instability at relatively 
low pressure, $CeIn_3$ is also thought to have a strongly first order 
$\alpha-\gamma$ transition at high pressures\cite{CEIN3}.  
Superconductivity is only observed in a narrow 
range of pressure and temperature around the antiferromagnetic 
quantum critical point. Because of the wide separation 
in pressure between the magnetic and density transitions and the 
strongly first order nature of the latter, one would expect the 
observed superconductivity to be magnetically mediated.  In the
tetragonal compounds $CeMIn_5$, however, superconductivity is 
observed over a wide range of pressures.  Were an $\alpha-\gamma$ 
transition present in these compounds, the critical end point would 
be expected to be at much lower temperatures than in $CeIn_3$ due 
to the role of anisotropy as discussed above.  This would result in 
stronger density fluctuations in the neighborhood of 
the $\alpha-\gamma$ instability.  Could this be another example 
where antiferromagnetic and low ${\bf q_0}$ density fluctuations 
both contribute to the attractive pairing interaction in the  
$d_{x^2-y^2}$ Cooper state?  This would explain the unusually wide 
extent of the superconducting domes observed in these materials.

It would not be surprising that such a density transition has not been reported
because we expect its signature to be weak.  Moreover, it is likely to be
observable as a well-defined transition only over a very narrow range in
pressure in the temperature-pressure phase diagram and would require very 
careful examination pressure scans at fixed temperatures in order to 
detect it\cite{VolChange}.

\section{Outlook}

One can expect that the total effective interaction between particles in a
strongly correlated electron system to be very complex.  The interaction
will clearly depend on the charge, but also more generally on the spin and
current carried by the particles.  The border of a density or spin or
current instability is characterized by strongly enhanced order-parameter
fluctuations and it is therefore plausible that the dominant interaction
channel is mediated by the density, spin or current fluctuations,
respectively.

In this paper we have shown how the framework developed for systems on the
border of magnetism can be translated to describe systems on the border of
density instabilities.  A striking feature of the model we have considered
is that the most robust pairing is obtained in the spin-singlet 
$d_{x^2-y^2}$ Cooper state on the border of both the density and spin 
instabilities.  However, crucially the wavevector ${\bf q_0}$ at which 
the response function is most enhanced is different in the two cases.  
Density fluctuations give rise to the highest superconducting $T_c$ 
for ${\bf q_0}$ near the center of the Brillouin zone while magnetic 
pairing is strongest for ${\bf q_0} = [\pi,\pi]$.  While it is 
possible to construct a Cooper pair state that samples mainly the most 
attractive regions of the density and magnetic interaction for 
${\bf q_0} = [\pi,\pi]$, the attraction is weaker in the density channel 
because the minimum separation of the two interacting particles is larger 
in the $d_{xy}$ state for the density interaction than in the $d_{x^2-y^2}$ 
state for the magnetic interaction (Fig. 1a). For low ${\bf q_0}$, however, 
the density interaction is mostly attractive provided that the particles 
are not on the same site (Fig. 4a) and thus the most favored state 
is $d_{x^2-y^2}$ in which the two interacting particles can take advantage
of the strong nearest-neighbor attraction.

It would seem that the same argument could apply to the low 
${\bf q_0}$ magnetic interaction.  However, in contrast to the 
density interaction which has the same sign in the spin-singlet 
and spin-triplet channels, the magnetic interaction depends on the 
relative spin orientation of the two interacting particles
and thus has a different sign for the two cases.  Magnetic pairing 
in the spin-singlet state is only possible if the real space 
interaction has sufficiently short-wavelength oscillations.  
Therefore, when ${\bf q_0}$ is near the center of the Brillouin zone 
magnetic pairing in the spin-singlet state is not possible, but is 
allowed in principle in the spin-triplet state for which the magnetic 
interaction has the opposite sign.  However, magnetic pairing in this 
state has the disadvantage that only the exchange of 
spin fluctuations polarized along the direction of the interacting 
spins, i.e., the longitudinal fluctuations contribute to the
particle interaction.  For a spin rotationally invariant system, 
both the longitudinal and transverse spin fluctuations contribute to 
pairing only for a spin-singlet Cooper state.  For the model considered 
in Refs\cite{ML1,ML2,ML3} this effect results in much weaker pairing 
on the border of ferromagnetism (${\bf q_0} = 0$) than 
antiferromagnetism with ${\bf q_0} = [\pi,\pi]$.

Another disadvantage of being on the border of ferromagnetism is that 
for otherwise similar conditions the suppression of $T_c$ due to the 
self-interaction arising from the exchange of magnetic fluctuations 
is stronger than in the corresponding case on the border of antiferromagnetism. 
This disadvantage can be mitigated in systems with strong magnetic anisotropy 
in that the effect of the transverse magnetic fluctuations on the 
self-interaction would be suppressed while the strength of the pairing 
interaction arising from the longitudinal magnetic fluctuations need not be 
reduced.  This may apply in systems with strong spin-orbit interactions or 
in the spin-polarized state close to the border of ferromagnetism.

These arguments ~\cite{ML1,ML2} have stimulated a new search for evidence of 
superconductivity on the border of itinerant electron ferromagnetism in 
cases where spin anisotropy is expected to be pronounced, such as $UGe_2$. 
This search has proved fruitful because it led to the first observation of 
the coexistence of superconductivity and itinerant electron ferromagnetism 
in $UGe_2$\cite{Saxena} and shortly thereafter in $ZrZn_2$\cite{Pfleiderer} 
and $URhGe$\cite{Huxley}.

In the previous section and in Figs. 10 and 11 we gave simple
arguments to explain how the pairing effect of the interactions are
strengthened by a tetragonal distortion in our model.  However, the same
effects also contribute to an enhanced self-interaction which acts to
suppress $T_c$.  The relative importance of the pair-forming and 
pair-breaking effects of the effective interaction cannot be inferred 
solely from the above physical picture for the density channel and 
the analogous arguments given in Ref.\cite{ML3} for the magnetic channel.  
The numerical calculations show that for most cases considered here 
and in Ref.\cite{ML3} the pair-forming effects dominate.

A most striking manifestation of the interplay between the pair-forming and
pair-breaking tendency of both the density and magnetic interactions is the
breakdown of the McMillan-style expression for $T_c$.  This was noted in 
Refs.\cite{ML1,ML2,ML3} and has been interpreted in Ref.\cite{Abanov}
in terms of the important role played by the incoherent 
part of the Green function which is ignored in the simplest treatments, 
but is included in the present and earlier work where the full 
momentum and frequency dependence of the self-energy is taken
into account.

In this and our earlier work we deliberately avoided modelling a specific
system since our main goal is to gain insights into the nature of the
pairing problem on the border of a density and spin instabilities.  We have
focussed on understanding trends and certain general factors affecting the
robustness of the pairing mechanism.  Even the simplest models considered
display surprising sensitivity to factors such as the nature of the
instability, the wavevector ${\bf q_0}$ at which it occurs, the total 
spin of the Cooper pair, details of the electronic and lattice structure 
as well as the form of the relevant response function.  Therefore, one 
should exercise caution in making quantitative comparisons between 
the results of our calculations and experiment.

In particular, our model may not apply to situations where there is a large,
local in space, contribution to the dynamical response function.  This would
not contribute to the pairing interaction for anisotropic Cooper states, but
could greatly enhance the self-interaction effect that is pair-breaking.
This could for instance greatly increase the sensitivity of $T_c$ to lattice
anisotropy as observed in $CeMIn_5$ systems and to the correlation length 
($1/\kappa$) characterising the relevant response function as indirectly 
seen in the strong pressure dependence of $T_c$ in, for example, $CeIn_3$.  
Such a local contribution to the magnetic response function has been 
observed in heavy fermion systems\cite{Local}.

The results of the calculations would be very sensitive to the particular
choice of the wavevector dependence of the response function.  In cases
where it falls off in ${\bf q}$ faster than in our model, the response is
appreciably enhanced in a smaller portion of the Brillouin zone and one
would then expect the effect of the density or magnetic interaction on the
thermal, transport and superconducting properties to be reduced.  This could
explain the surprisingly weak effects on these properties of the CDW
fluctuations in systems such as $NbSe_2$\cite{NbSe2}.

At first sight, our results seem to imply that anisotropic forms of
superconductivity should be a generic property of systems on the border of
density and magnetic instabilities.  It may seem surprising therefore that
there are still relatively few observations of this phenomenon.  In 
addition to the sensitivity of $T_c$ to details of the system as 
discussed above, in many cases the multiplicity of bands and lattice 
structure may be unfavorable for pairing to such an extent that 
quenched disorder may completely suppress superconductivity.  An 
illustration of this latter point is the dramatic collapse of the spin-triplet 
superconducting $T_c$ in $Sr_2RuO_4$ in the presence of Al impurity 
concentrations as low as 0.1\% \cite{SrRuO4}.  Another factor that
may explain the absence of superconductivity is the common occurrence of
first order rather than continuous magnetic as well as density 
instabilities.  Our results show that in many cases one has to be 
close to the instability.  A first order transition may make this
region of the phase diagram inaccessible.

The results of the mean-field calculations presented here and in our earlier
papers show that robust pairing can occur in both density and magnetic
channels under suitable conditions.  Therefore, it would seem that one
should not favor one mechanism over another in the search for new examples
of high temperature superconductivity.  This conclusion may turn out to be
incorrect when corrections to the one-loop mean-field calculations are taken
into account.  In contrast to the conventional electron-phonon pairing
theory where corrections to the Eliashberg value of $T_c$ are small, it has
been argued for many years\cite{BealMonod} that this may not be the case 
for other types of pairing mechanisms.

It has been shown that the mean-field approximation of the kind we have
considered here qualitatively breaks down in a half-filled Hubbard model in
2D which is a Mott insulator with long-range antiferromagnetic order at
absolute zero.  This breakdown has been interpreted in 
terms of the effect of thermal magnetic fluctuations in the renormalized 
classical regime\cite{Tremblay}.  Thermal density fluctuations near 
Peirels CDW transition in 2D also lead to qualitative changes to the 
electronic spectrum that are not captured in the present model\cite{MS}.

Recent non-perturbative calculations have shown that dynamical fluctuations
even at the Gaussian level are sufficient to cause a breakdown of the
present mean-field model for sufficiently small $\kappa^2$\cite{Monthoux}.
In this and our earlier work on the magnetic interaction model, we had to
solve the Eliashberg equations for the superconducting transition temperature
$T_c$ for very many choices of model parameters. Even with the best numerical
algorithms, this is only practically feasible, especially in 3D, if the overall
scale of the $T_c$'s obtained is rather high (say of the order of 0.1t). 
Our choice of the charactersitic density-fluctuation temperature 
$T_{DF} = 2t/3$ (or spin-flucutation temperature $T_{SF} = 2t/3$ 
in our earlier work) was therefore dictated by such considerations.
We now know\cite{Monthoux} that for these values of $T_{DF}$ (or $T_{SF}$)
and the range of other model parameters considered here and in our 
earlier papers that vertex corrections are important. Our results are
nevertheless useful if interpreted in the following way. We expect, and 
have checked in a number of cases\cite{ML1}, that the trends in $T_c/T_{DF}$
(or $T_c/T_{SF}$) are weakly dependent on the abolute scale of the 
characteristic temperatures $T_{DF}$ and $T_{SF}$. Therefore the conclusions
drawn from our calculations are expected to remain qualitatively correct
for smaller values of $T_{DF}$ or $T_{SF}$ and hence $T_c$, values for which
the mean-field theory of superconductivity is likely to be more accurate.
  
It has generally been believed that the most important factor in determining
the applicability of Migdal's theorem is the form and parameters entering
the relevant fluctuation spectrum.  Therefore, a surprising finding was that
in the range of model parameters studied in Ref.\cite{Monthoux}, the vertex 
corrections to the Eliashberg self-energy led to qualitatively different 
electron spectral functions for a coupling to magnetic and density 
fluctuations with identical fluctuation spectra.  In those calculations, 
it was found that the corrections to the Eliashberg theory enhanced the 
magnetic interaction, but suppressed the density interaction.  This effect 
can readily be seen at the two-loop level.

The contribution of the transverse magnetic fluctuations to the first order
vertex correction is opposite in sign to that of longitudinal ones and 
dominates.  On the other hand, in the density channel one has essentially 
only 'longitudinal' fluctuations, which as in the magnetic case suppress 
the interaction at this level of approximation.  It is also known that 
the two fluctuation exchange processes lead to the enhancement of the 
spin-singlet magnetic pairing interaction\cite{Vertex}. While detailed 
calculations of $T_c$ beyond the single-fluctuation exchange approximation 
have yet to be carried out, the above findings suggest that spin-singlet 
magnetic pairing may turn out to be more robust than density pairing under 
otherwise equivalent conditions.

\section{Acknowledgments}

We would like to thank A.V. Chubukov, P. Coleman, F.M. Grosche, S.R. Julian, 
P.B. Littlewood, A.J. Millis, A.P. Mackenzie, D. Pines, D.J. Scalapino,  
M. Sigrist and H. Yuan, for discussions on this and related topics. We 
acknowledge the support of the EPSRC, the Newton Trust and the Royal Society.

\section{Appendix}

We consider quasiparticles on a cubic or tetragonal lattice. 
We assume that the dominant interaction is in the density
channel and postulate the following low-energy effective action 
for the quasiparticles:

\begin{eqnarray}
S_{eff} & = & \sum_{{\bf p},\alpha}\int_0^\beta d\tau 
\psi^\dagger_{{\bf p},\alpha}(\tau)\Big(\partial_\tau + 
\epsilon_{\bf p} - \mu\Big) \psi_{{\bf p},\alpha}(\tau)
+  {I\over N}\sum_{\bf q}\int_0^\beta d\tau
\rho_{\uparrow}({\bf q},\tau)
\rho_{\downarrow}(-{\bf q},\tau)\nonumber\\
& & - {g^2\over 2 N}\sum_{\bf q}\int_0^\beta 
d\tau \int_0^\beta d\tau' \chi({\bf q},\tau-\tau')
\rho({\bf q},\tau)\rho(-{\bf q},\tau')
\label{Seff}
\end{eqnarray}

\noindent where $N$ is the number of allowed wavevectors in the Brillouin
zone and the carrier density $\rho_{\sigma}({\bf q},\tau)$ is given by

\begin{equation}
\rho_\sigma({\bf q},\tau) \equiv \sum_{{\bf p}} 
\psi^\dagger_{{\bf p} + {\bf q},\sigma}(\tau)
\psi_{{\bf p},\sigma}(\tau)
\label{Rho}
\end{equation}

\noindent and $\rho({\bf q},\tau) = \sum_{\sigma}\rho_\sigma({\bf q},\tau)$.
The quasiparticle dispersion relation $\epsilon_{\bf p}$ is defined in 
Eq.~(\ref{eps}), $\mu$ denotes the chemical potential, $\beta$ the 
inverse temperature, $g^2$ the coupling constant and 
$\psi^\dagger_{{\bf p},\sigma}$ and $\psi_{{\bf p},\sigma}$ are
Grassmann variables. We measure temperatures, 
frequencies and energies in the same units. Our effective density
interaction consists of an induced part, the last term in Eq.~(\ref{Seff}),
and a local on-site Coulomb repulsion, the second term in Eq.~(\ref{Seff}).

The retarded generalized susceptibility $\chi({\bf q},\omega)$ 
that defines the effective interaction, Eq.~(\ref{Seff}), is defined
in Eq.~(\ref{chiML}).

The density-fluctuation propagator on the imaginary axis, 
$\chi({\bf q},i\nu_n)$ is related to the imaginary part of the 
response function $Im\chi({\bf q},\omega)$, Eq.~(\ref{chiML}), 
via the spectral representation

\begin{equation}
\chi({\bf q},i\nu_n) = -\int_{-\infty}^{+\infty}{d\omega\over \pi}
{Im\chi({\bf q},\omega)\over i\nu_n - \omega}
\label{chi_mats}
\end{equation}

\noindent To get $\chi({\bf q},i\nu_n)$ to decay as $1/\nu_n^2$ as
$\nu_n \rightarrow \infty$, as it should, we introduce a cutoff 
$\omega_0$ and take $Im\chi({\bf q},\omega) = 0$ for $\omega 
\geq \omega_0$. A natural choice for the cutoff is $\omega_0 
= \eta(\widehat{q})\kappa_0^2$. 

The Eliashberg equations for the critical 
temperature $T_c$ in the Matsubara representation reduce, for the
effective action Eq.~(\ref{Seff}), to

\begin{equation}
\Sigma({\bf p},i\omega_n) = g^2{T\over N}\sum_{\Omega_n}
\sum_{\bf k}\chi({\bf p}-{\bf k},i\omega_n-i\Omega_n)
G({\bf k},i\Omega_n)
\label{Sigma}
\end{equation}

\begin{equation}
G({\bf p},i\omega_n) = {1\over i\omega_n - (\epsilon_{\bf p}-\mu) 
- \Sigma({\bf p} ,i\omega_n)}
\label{Green}
\end{equation}

\begin{eqnarray}
\Lambda(T)\Phi({\bf p},i\omega_n) & = & 
{T\over N}\sum_{\Omega_n}\sum_{\bf k}
\Big(g^2\chi({\bf p} - {\bf k},i\omega_n - i\Omega_n) - I\Big)
|G({\bf k},i\Omega_n)|^2 \Phi({\bf k},i\Omega_n) \nonumber \\
\Lambda(T) & = & 1 \longrightarrow T = T_c
\label{Gap}
\end{eqnarray}

\noindent where $\Sigma({\bf p},i\omega_n)$ is the quasiparticle 
self-energy, $G({\bf p},i\omega_n)$ the one-particle Green's 
function and $\Phi({\bf p},i\omega_n)$ the anomalous self-energy.
The Hartree terms coming from the on-site Coulomb repulsion and
induced density interaction have been absorbed in the definition 
of the chemical potential, which is adjusted to give an electron 
density of $n = 1.1$. $N$ is the total number of allowed wavevectors 
in the Brillouin Zone. Eq.~(\ref{Gap}) has been written for 
spin-singlet Cooper pairs. In the spin-triplet channel, the sign
and coefficient of the term 
$g^2\chi({\bf p} - {\bf k},i\omega_n - i\Omega_n)$ 
remains unchanged. The on-site Coulomb interaction $I$ in 
Eq.~(\ref{Gap}) can be ignored since in the for spin-triplet 
Cooper pairs, the amplitude for the two particles to be on the
same site simultaneously vanishes.

The momentum convolutions in Eqs.~(\ref{Sigma},\ref{Gap}) are 
carried out with a Fast Fourier Transform algorithm on a 
$128 \times 128$ lattice for calculations in two dimensions and
$48 \times 48 \times 48$ lattice for three dimensional calculations. 
The frequency sums in both the  self-energy and linearized gap 
equations are treated with the renormalization group technique of 
Pao and Bickers\cite{PaoBickers}. We have kept between 8 and 16 
Matsubara frequencies at each stage of the renormalization procedure, 
starting with an initial temperature $T_0 = 0.4t$ in two dimensions
and $T_0 = 0.6t$ in three dimensions and cutoff $\Omega_c \approx 30t$.
The renormalization group acceleration technique restricts one to a 
discrete set of temperatures $T_0 > T_1 > T_2 \dots$. The critical 
temperature at which $\Lambda(T) = 1$ in Eq.~(\ref{Gap}) is 
determined by linear interpolation.

\begin{figure}
\centerline{\epsfysize=6.00in
\epsfbox{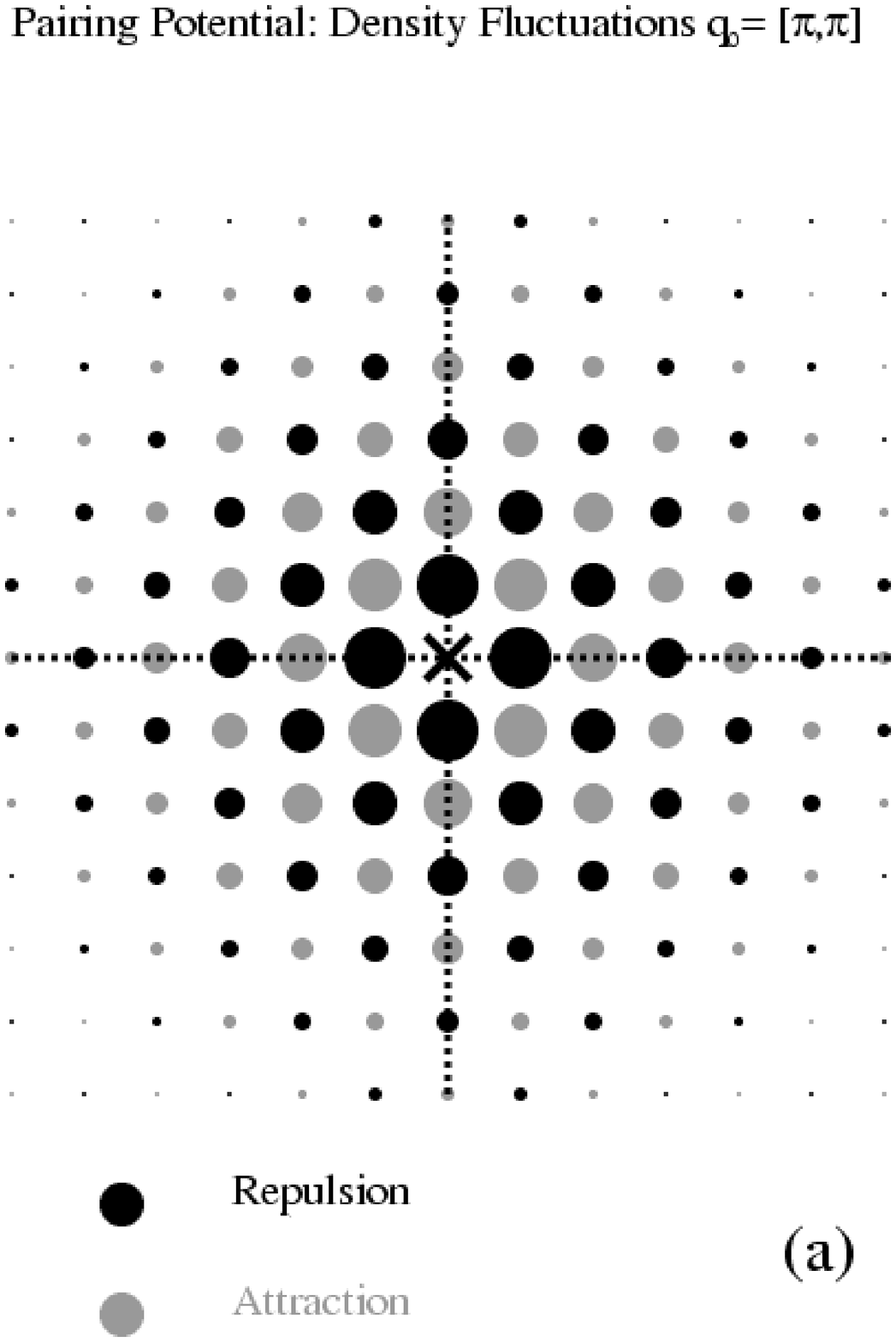}}
\centerline{\epsfysize=6.00in
\epsfbox{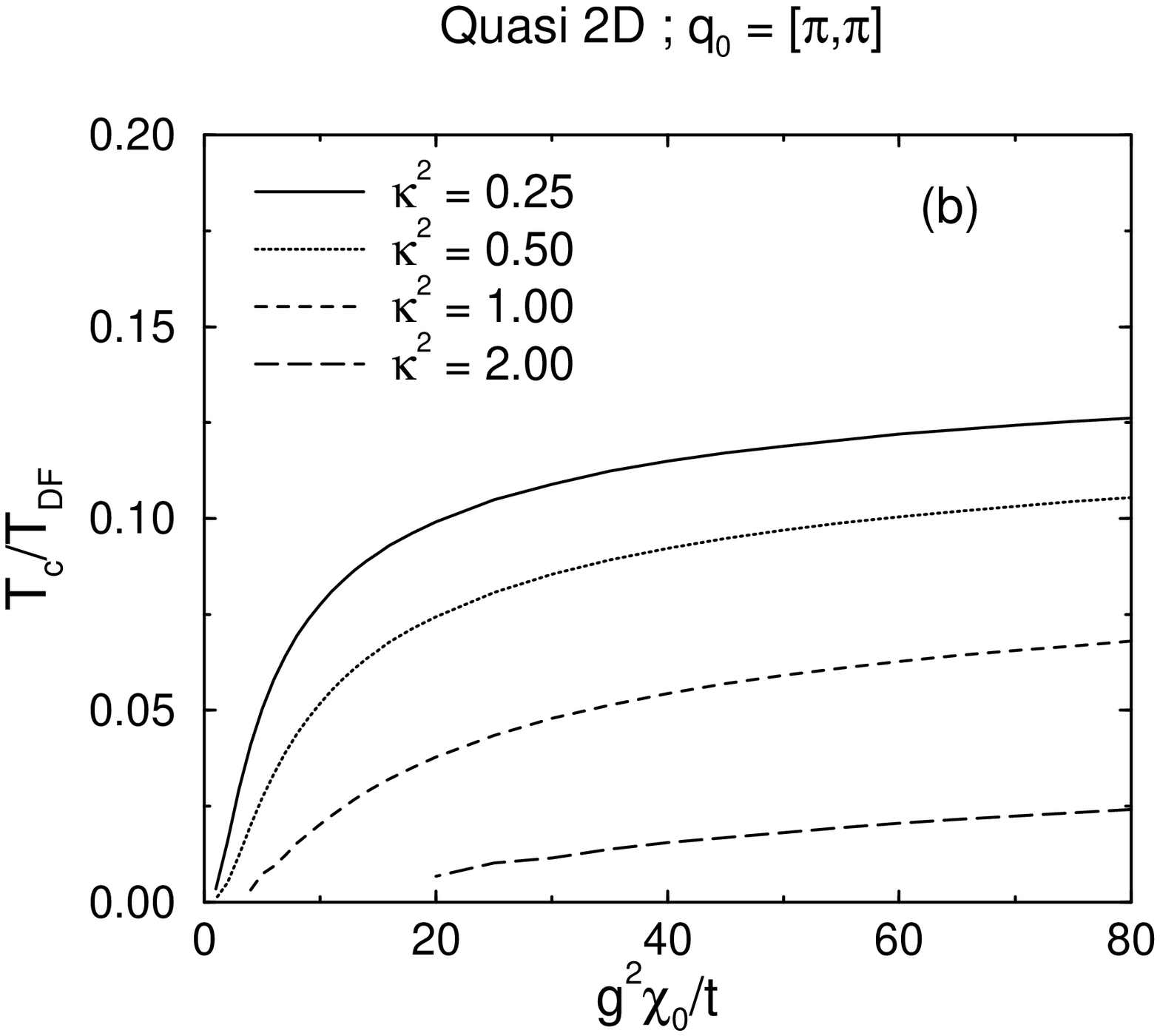}}
\centerline{\epsfysize=6.00in
\epsfbox{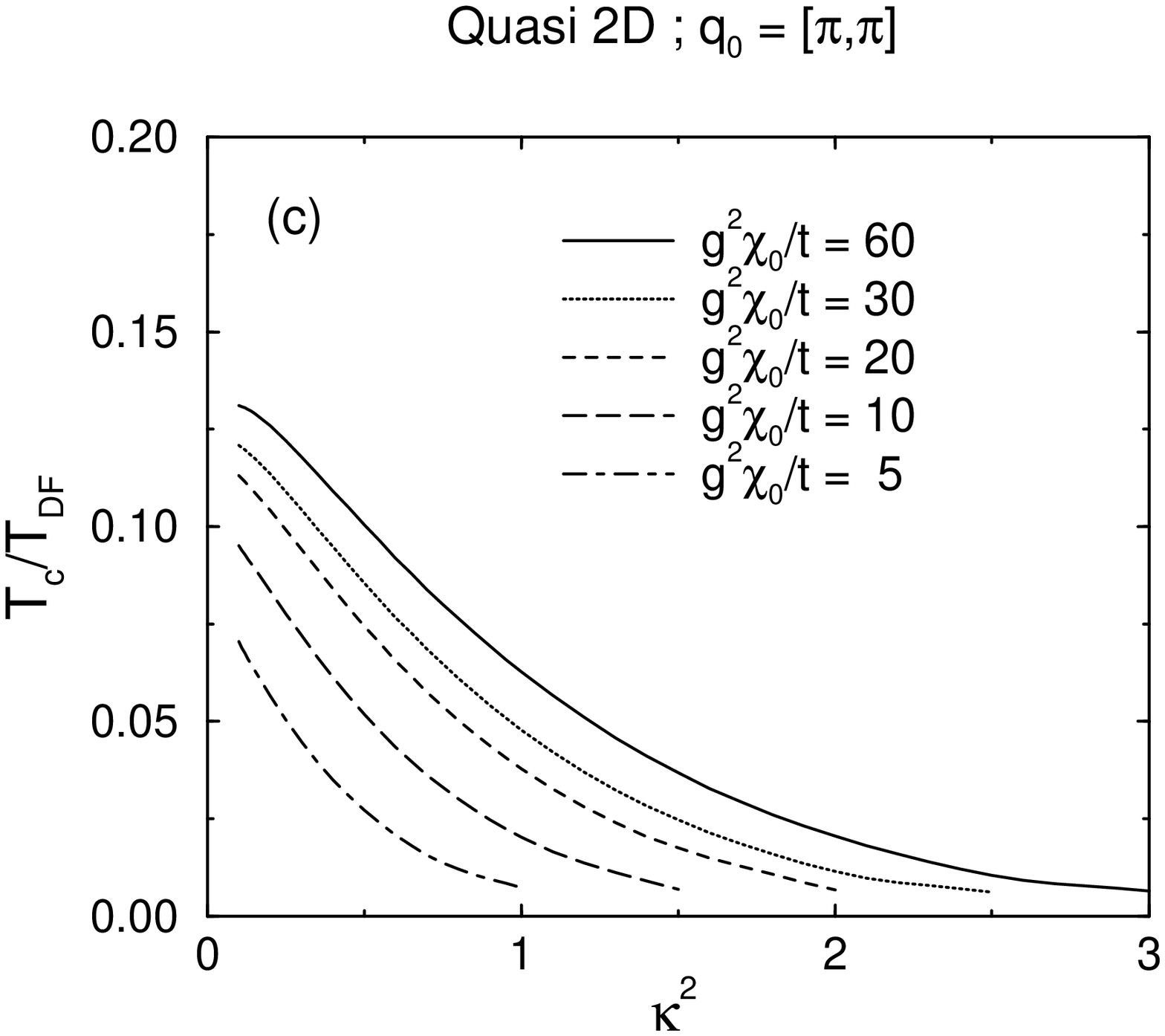}}
\vskip 1.0cm
\caption{(a) Static pairing potential seen by a quasiparticle in a 
square lattice given that the other quasiparticle is at the origin 
(marked by a cross) for an incipient ordering wavevector 
${\bf q_0} = [\pi,\pi]$. The sites are colored black if the interaction 
is repulsive and light gray if it is attractive. The size of the 
circles represents, on a logarithmic scale, the absolute value of the 
static pairing potential. The dashed line indicates the nodal lines 
of the $d_{xy}$ Cooper state.
(b) and (c) show the Eliashberg $T_c/T_{DF}$ for a quasi two-dimensional 
system as a function of the coupling constant $g^2\chi_0/t$ (b) 
and correlation wavevector $\kappa^2$ (c) for the choice $T_{DF} = 2t/3$ 
and $\kappa_0^2 = 12$. }
\label{fig1}
\end{figure}

\begin{figure}
\centerline{\epsfysize=6.00in
\epsfbox{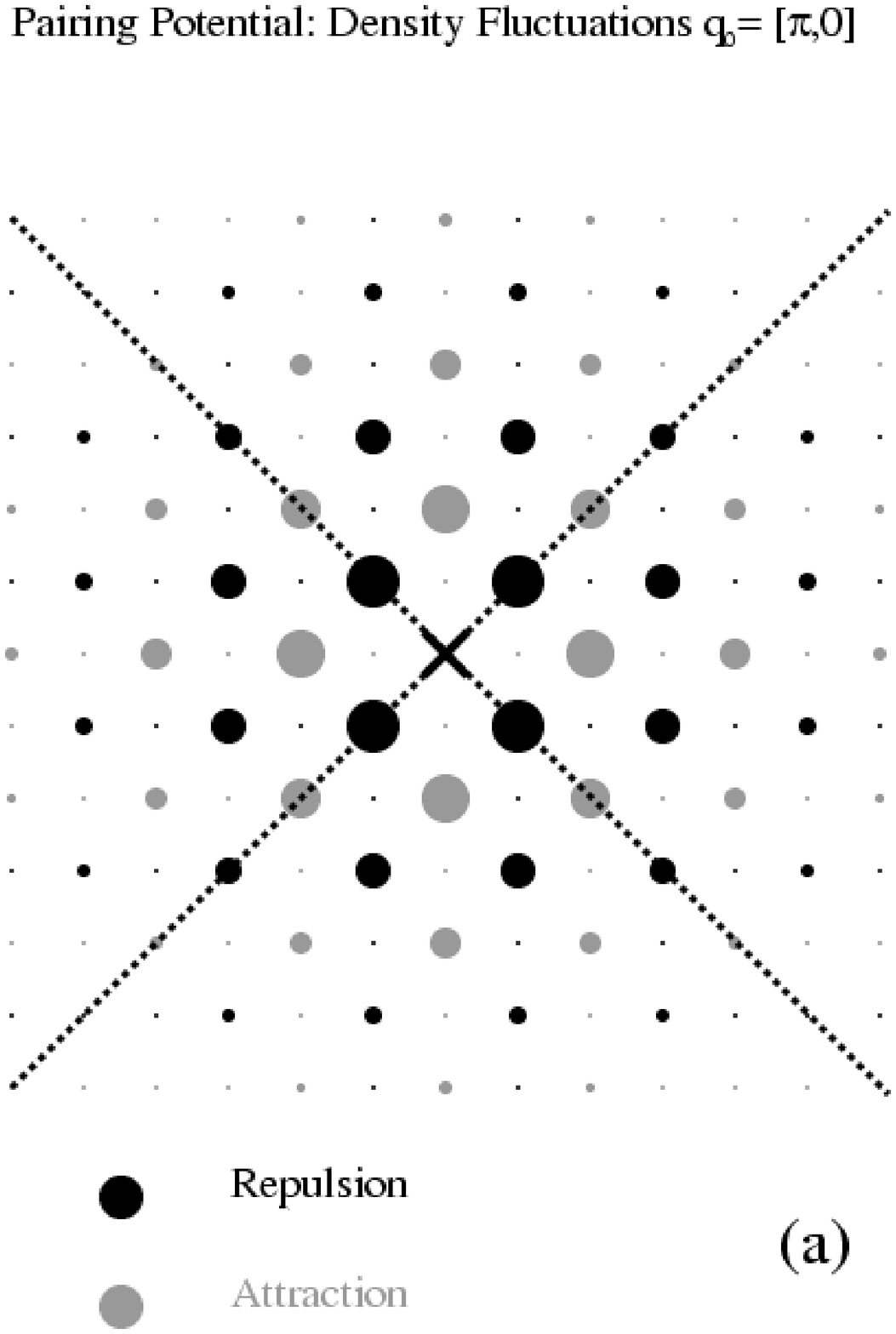}}
\centerline{\epsfysize=6.00in
\epsfbox{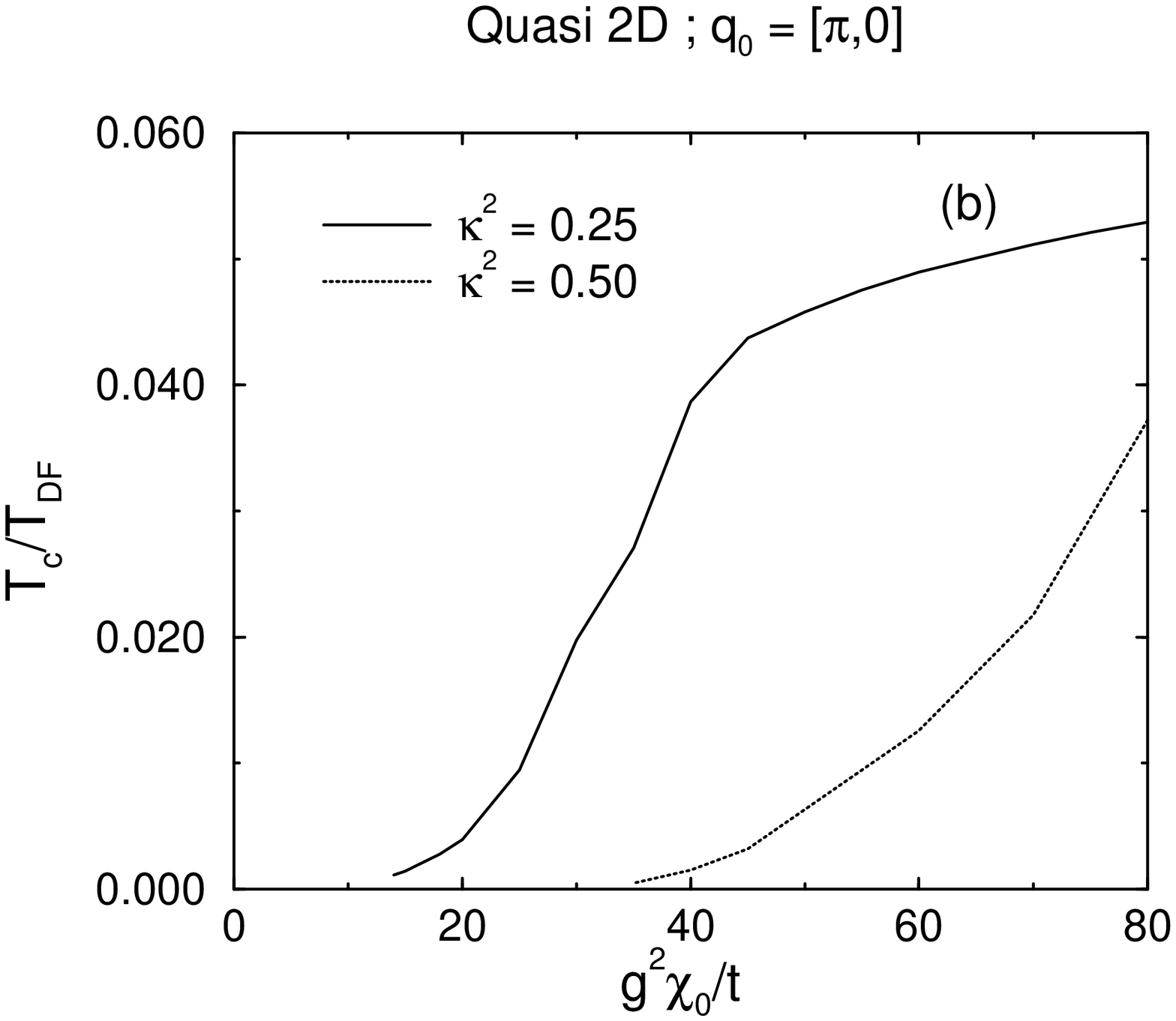}}
\centerline{\epsfysize=6.00in
\epsfbox{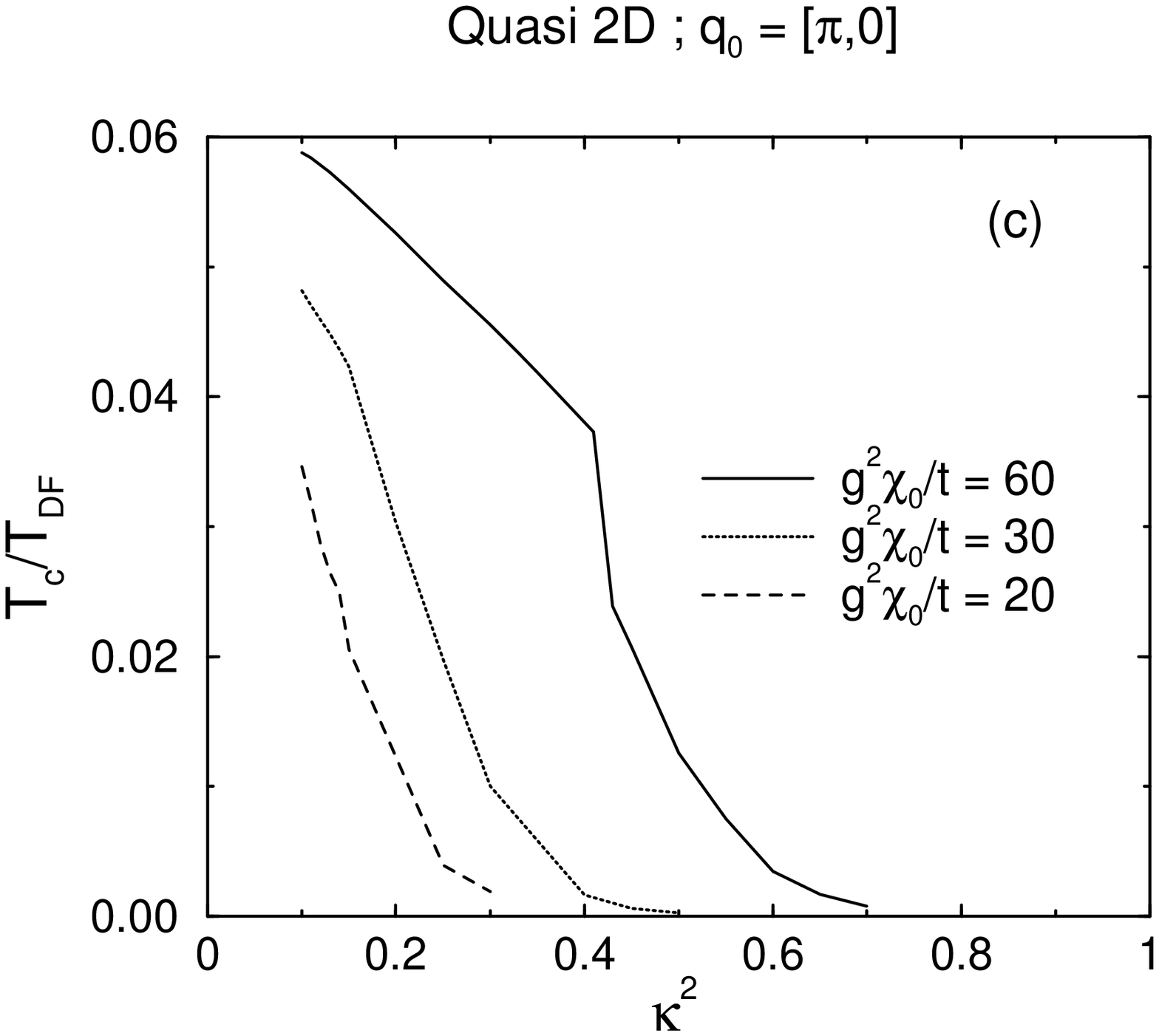}}
\vskip 1.0cm
\caption{(a) Static pairing potential seen by a quasiparticle in a 
square lattice given that the other quasiparticle is at the origin 
(marked by a cross) for an incipient ordering wavevector 
${\bf q_0} = [\pi,0]$. The sites are colored black if the interaction 
is repulsive and light gray if it is attractive. The size of the 
circles represents, on a logarithmic scale, the absolute value of the 
static pairing potential. The dashed line indicates the nodal lines 
of the $d_{x^2-y^2}$ Cooper state.
(b) and (c) show the Eliashberg $T_c/T_{DF}$ for a quasi two-dimensional 
system as a function of the coupling constant $g^2\chi_0/t$ (b) 
and correlation wavevector $\kappa^2$ (c) for the choice $T_{DF} = 2t/3$ 
and $\kappa_0^2 = 12$. }
\label{fig2}
\end{figure}

\begin{figure}
\centerline{\epsfysize=6.00in
\epsfbox{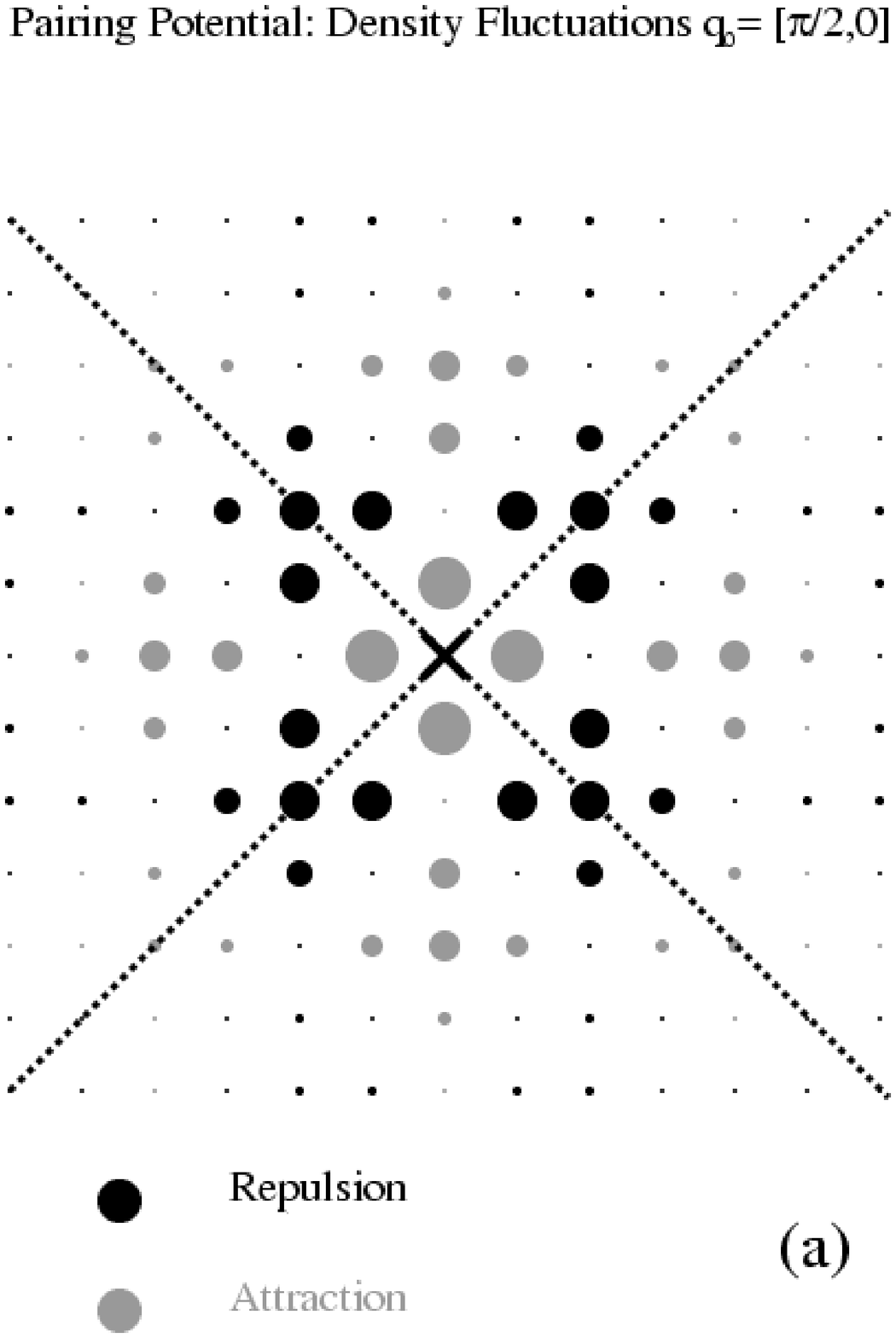}}
\centerline{\epsfysize=6.00in
\epsfbox{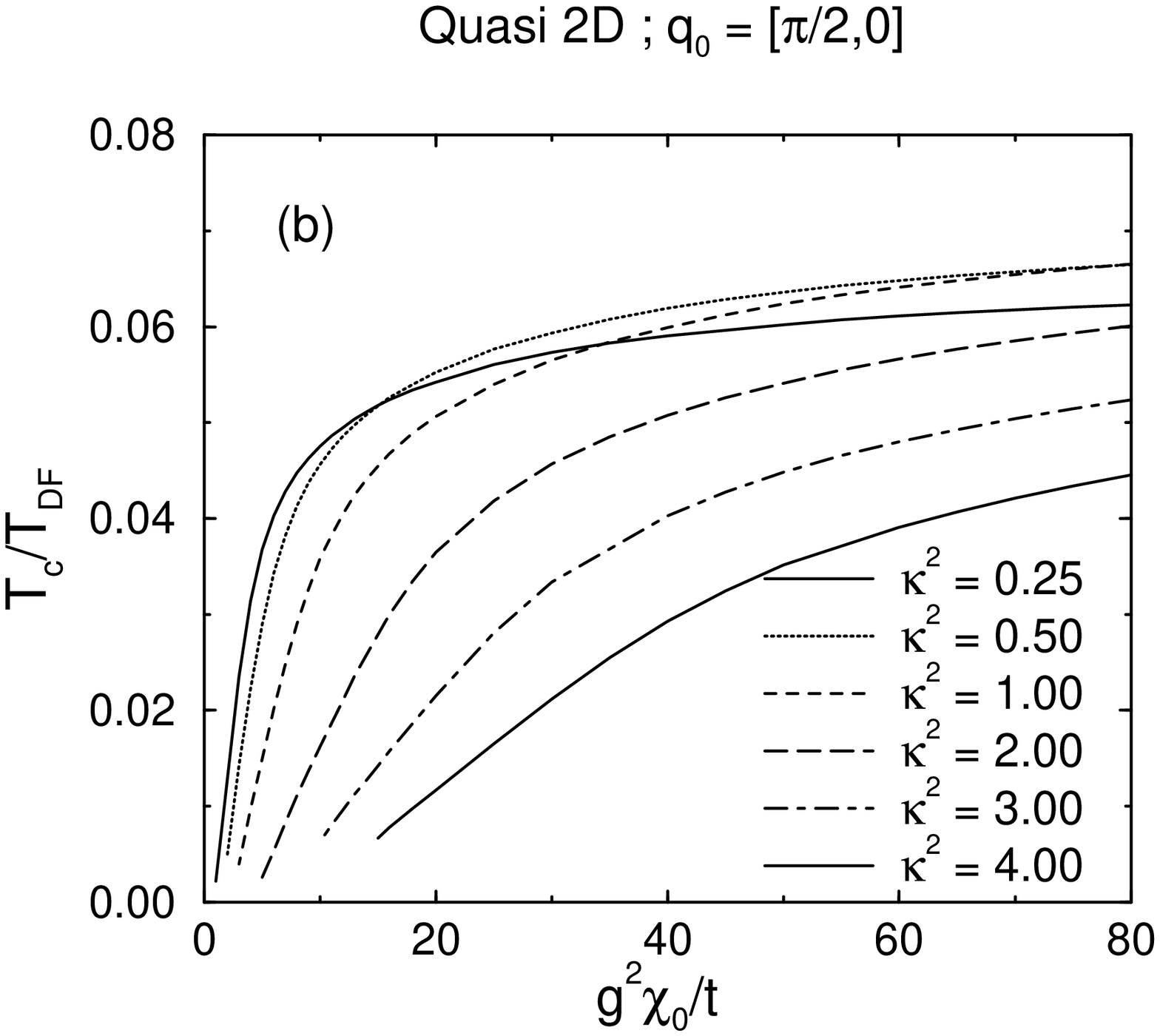}}
\centerline{\epsfysize=6.00in
\epsfbox{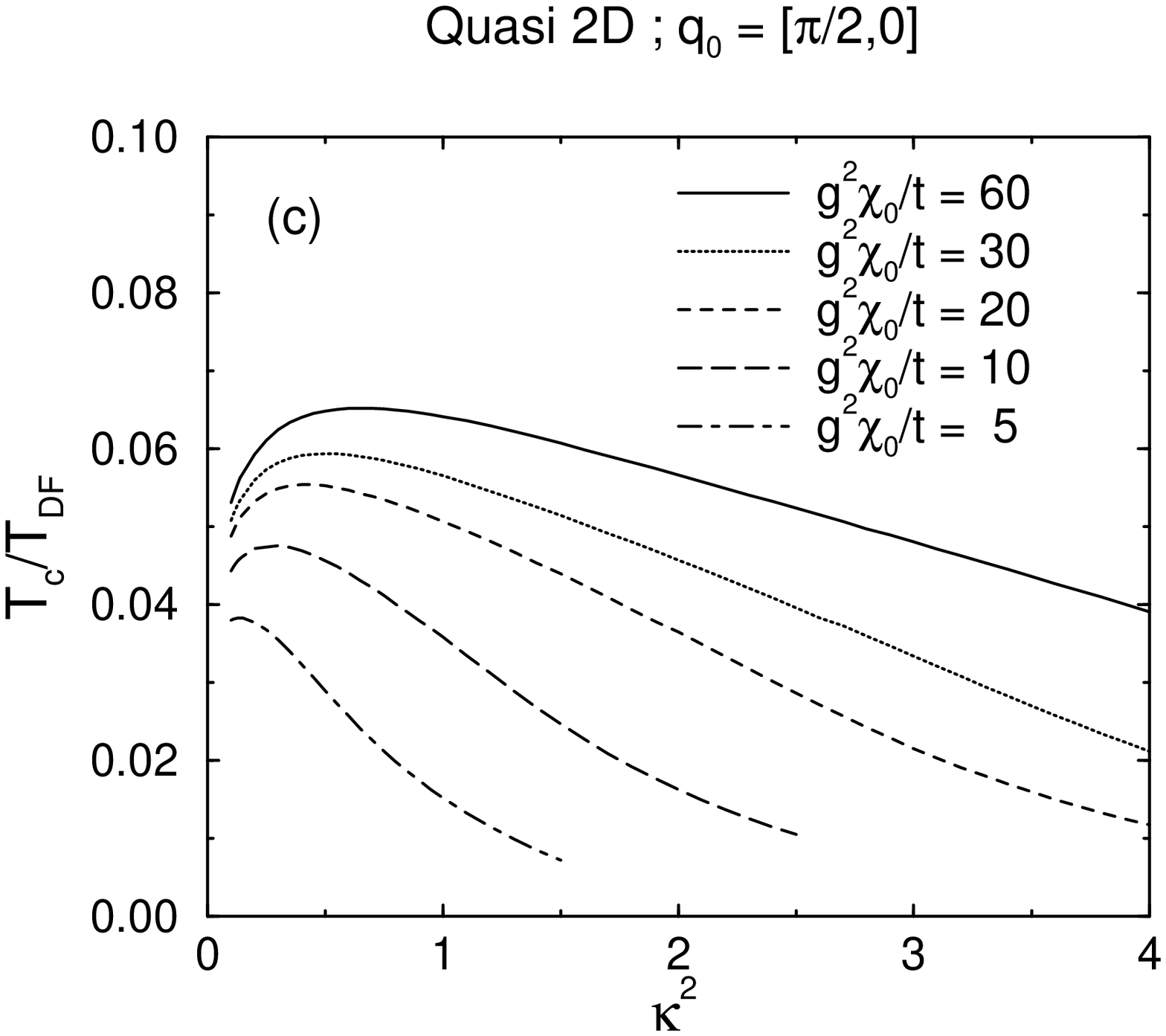}}
\vskip 1.0cm
\caption{(a) Static pairing potential seen by a quasiparticle in a 
square lattice given that the other quasiparticle is at the origin 
(marked by a cross) for an incipient ordering wavevector 
${\bf q_0} = [\pi/2,0]$. The sites are colored black if the interaction 
is repulsive and light gray if it is attractive. The size of the 
circles represents, on a logarithmic scale, the absolute value of the 
static pairing potential. The dashed line indicates the nodal lines 
of the $d_{x^2-y^2}$ Cooper state.
(b) and (c) show the Eliashberg $T_c/T_{DF}$ for a quasi two-dimensional 
system as a function of the coupling constant $g^2\chi_0/t$ (b) 
and correlation wavevector $\kappa^2$ (c) for the choice $T_{DF} = 2t/3$ 
and $\kappa_0^2 = 12$. }
\label{fig3}
\end{figure}

\begin{figure}
\centerline{\epsfysize=6.00in
\epsfbox{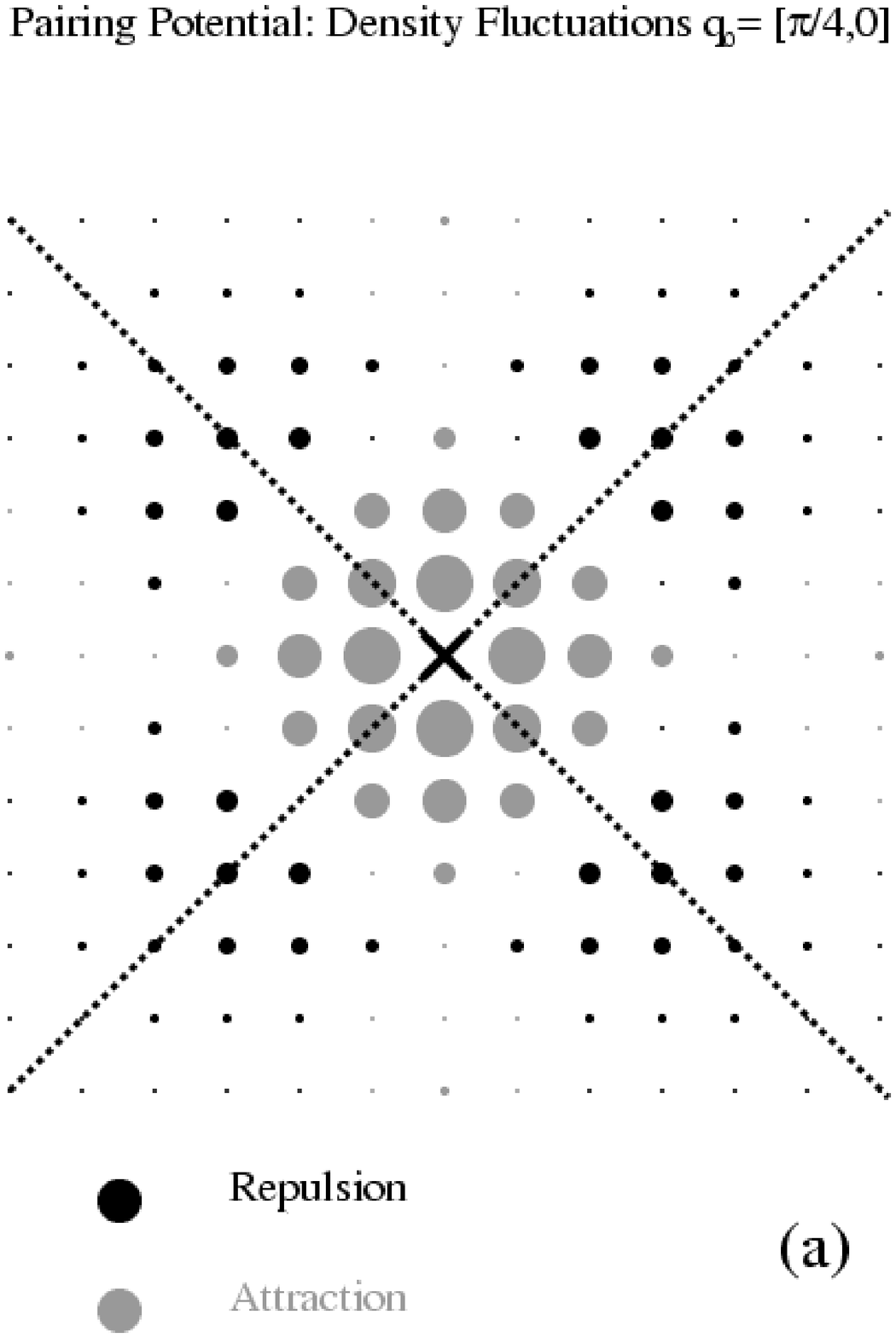}}
\centerline{\epsfysize=6.00in
\epsfbox{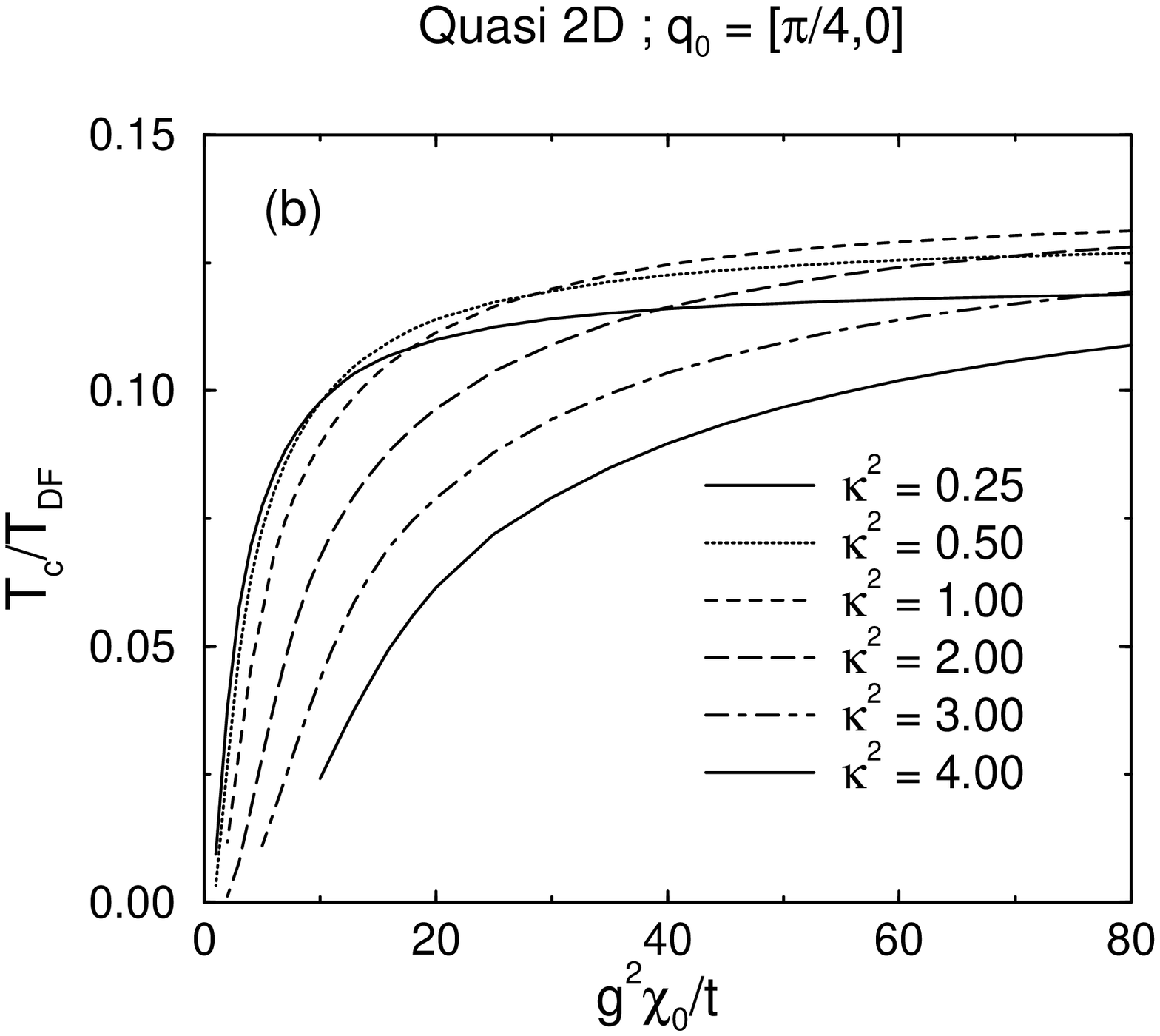}}
\centerline{\epsfysize=6.00in
\epsfbox{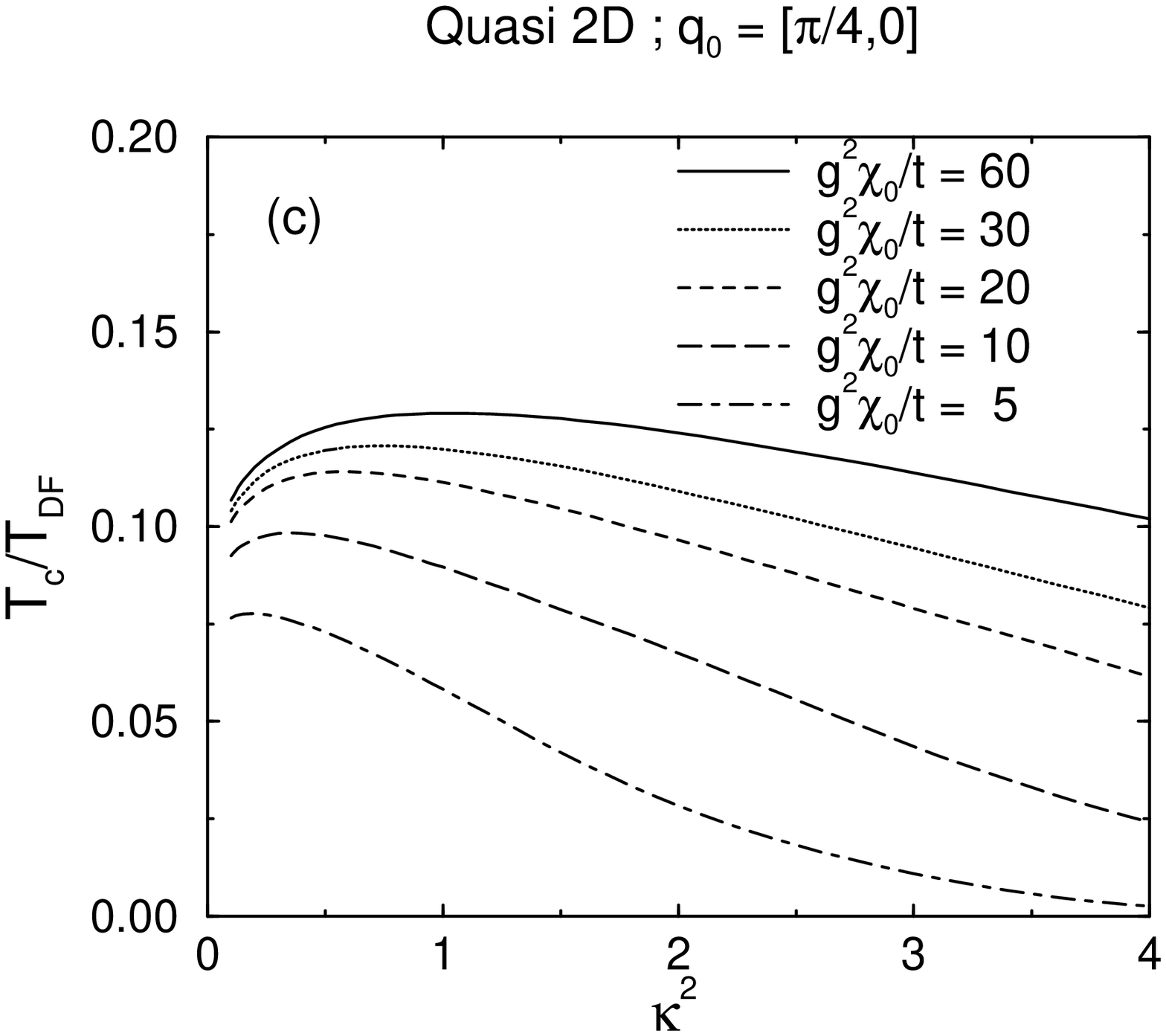}}
\vskip 1.0cm
\caption{(a) Static pairing potential seen by a quasiparticle in a 
square lattice given that the other quasiparticle is at the origin 
(marked by a cross) for an incipient ordering wavevector 
${\bf q_0} = [\pi/4,0]$. The sites are colored black if the interaction 
is repulsive and light gray if it is attractive. The size of the 
circles represents, on a logarithmic scale, the absolute value of the 
static pairing potential. The dashed line indicates the nodal lines 
of the $d_{x^2-y^2}$ Cooper state.
(b) and (c) show the Eliashberg $T_c/T_{DF}$ for a quasi two-dimensional 
system as a function of the coupling constant $g^2\chi_0/t$ (b) 
and correlation wavevector $\kappa^2$ (c) for the choice $T_{DF} = 2t/3$ 
and $\kappa_0^2 = 12$. }
\label{fig4}
\end{figure}

\begin{figure}
\centerline{\epsfysize=6.00in
\epsfbox{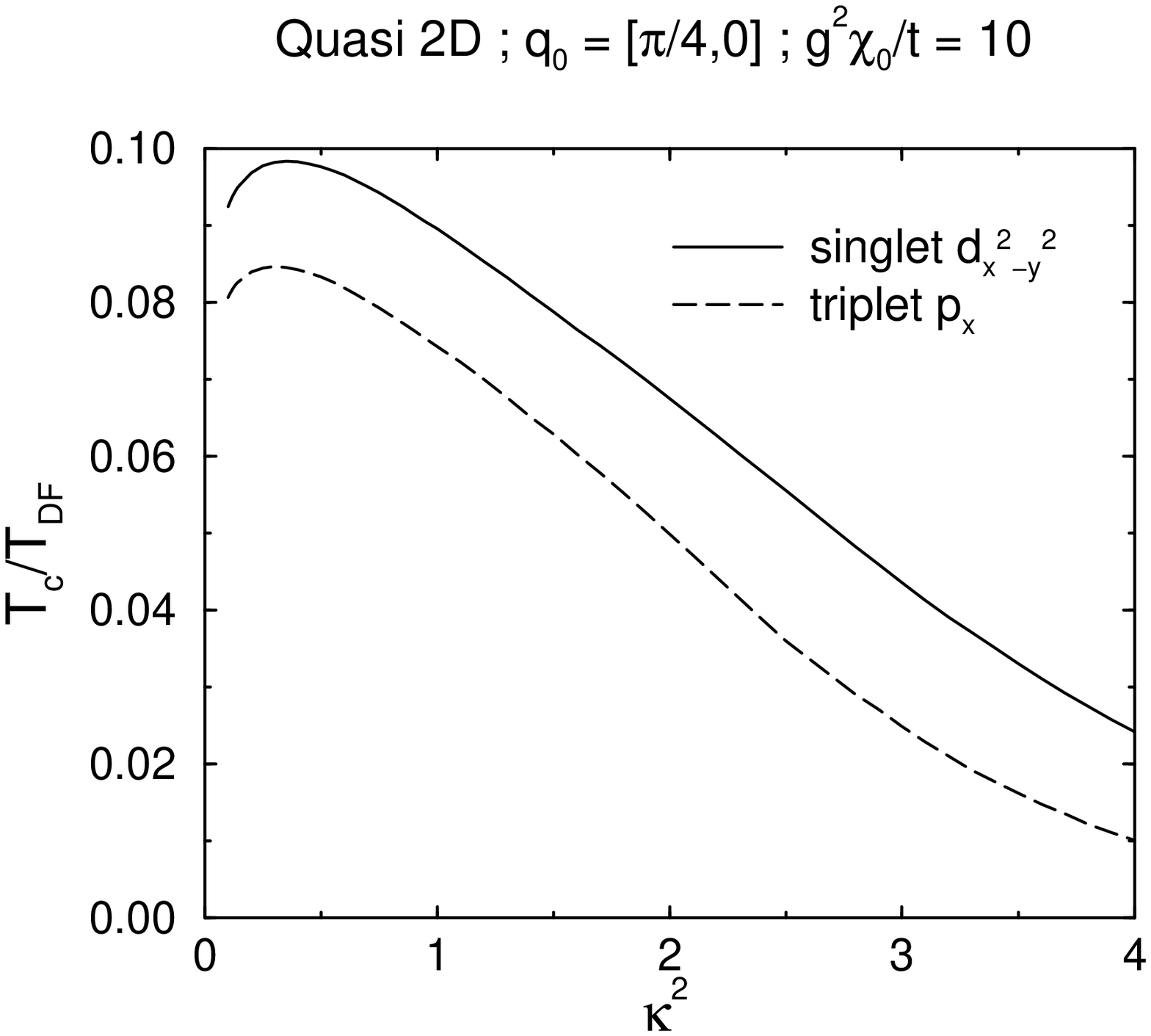}}
\vskip 1.0cm
\caption{Comparison of the Eliahsberg $T_c/T_{DF}$ for a quasi 
two-dimensional system with incipient ordering wavevector
${\bf q_0} = [\pi/4,0]$ in the spin-singlet $d_{x^2-y^2}$ versus
spin-triplet $p_x$ Cooper state. The model parameters
used in the calculations are $g^2\chi_0/t = 10$, $T_{DF} = 2t/3$
and $\kappa_0^2 = 12$. }
\label{fig5}
\end{figure}

\begin{figure}
\centerline{\epsfysize=6.00in
\epsfbox{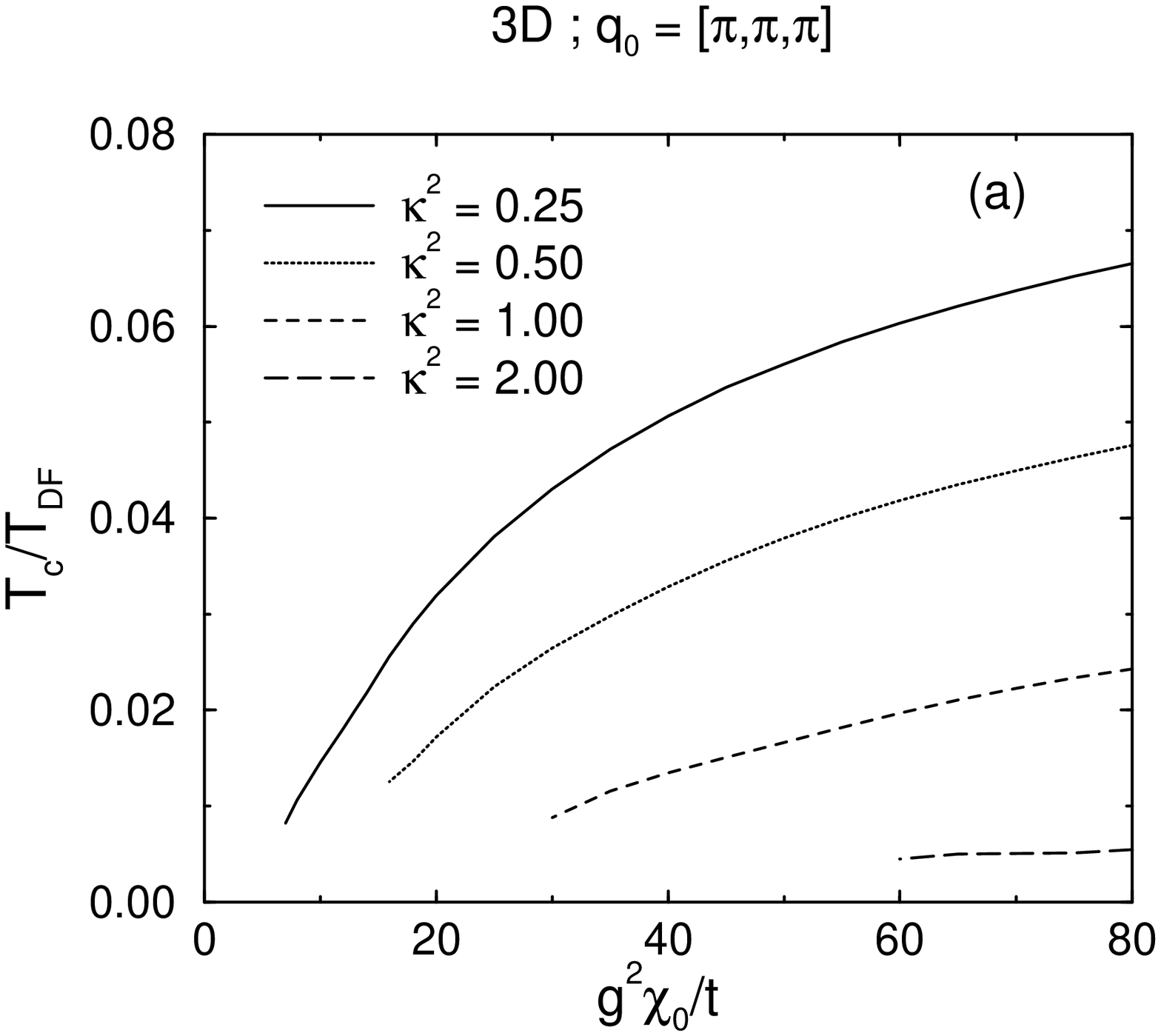}}
\centerline{\epsfysize=6.00in
\epsfbox{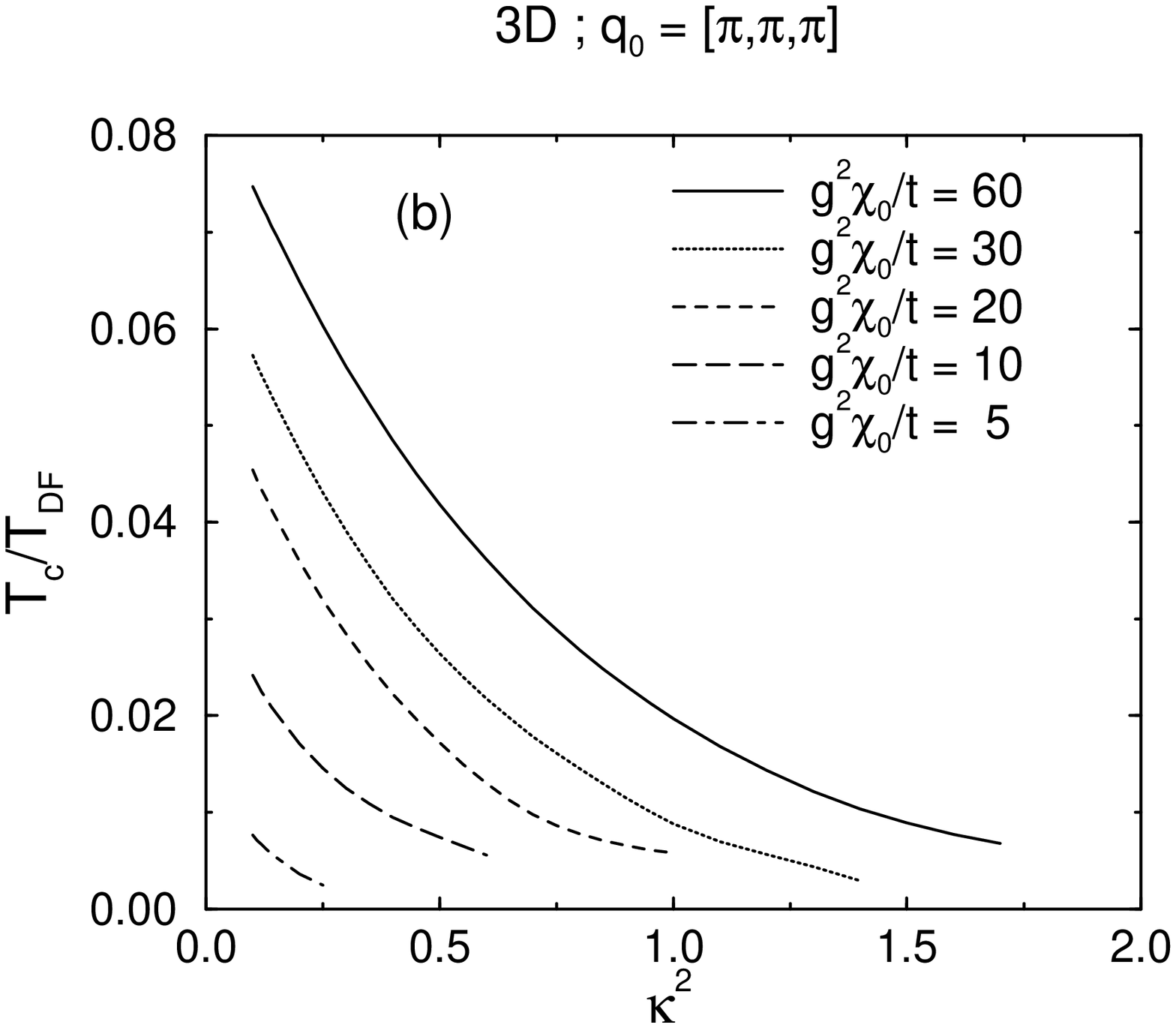}}
\vskip 1.0cm
\caption{(a) and (b) show the spin-singlet $d_{xy}$ Eliashberg $T_c/T_{DF}$ 
for a three-dimensional system with incipient ordering wavevector 
${\bf q_0} = [\pi,\pi,\pi]$ in as a function of the coupling constant 
$g^2\chi_0/t$ (a) and correlation wavevector $\kappa^2$ (b) for the 
choice $T_{DF} = 2t/3$ 
and $\kappa_0^2 = 12$. }
\label{fig6}
\end{figure}

\begin{figure}
\centerline{\epsfysize=6.00in
\epsfbox{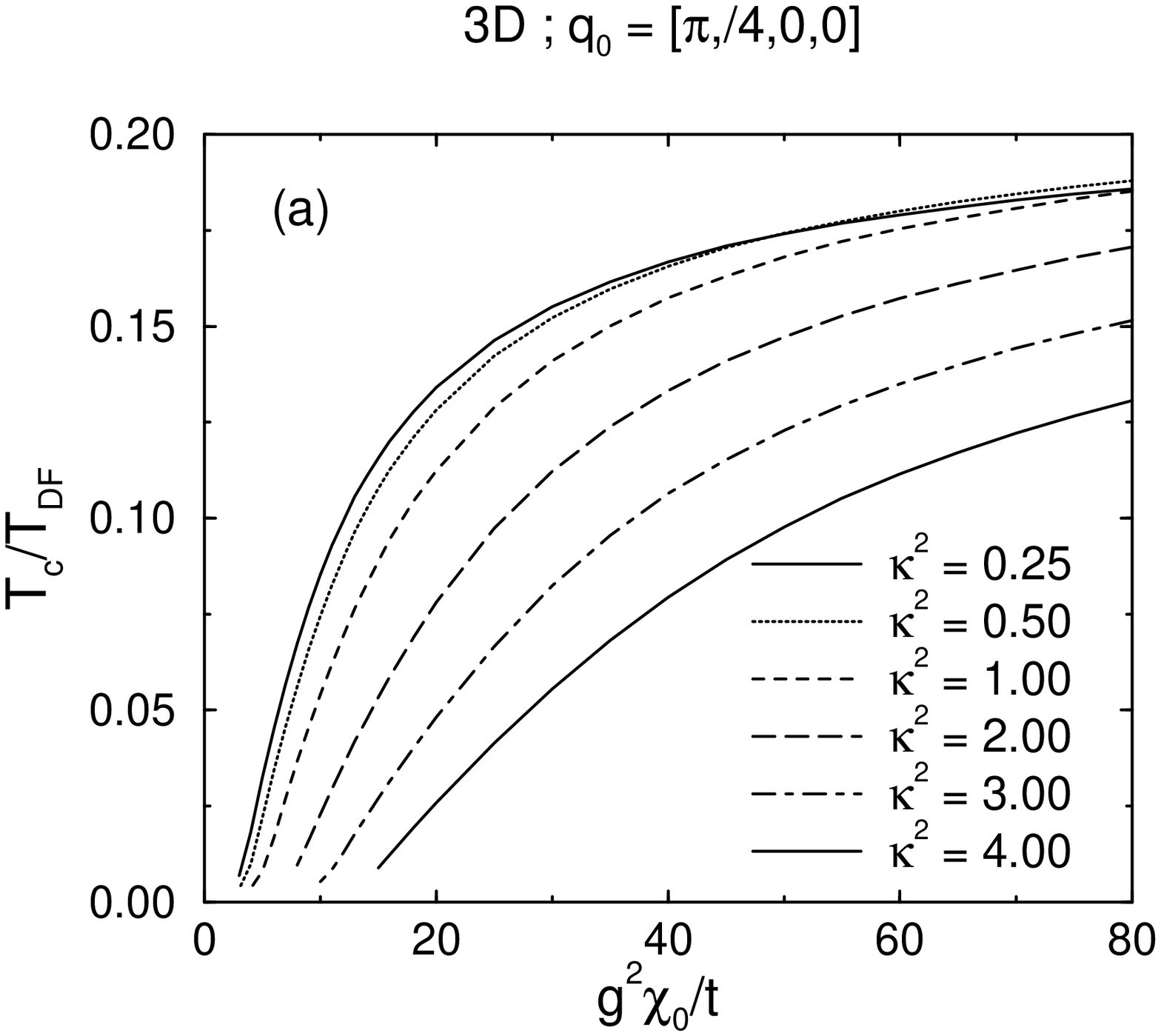}}
\centerline{\epsfysize=6.00in
\epsfbox{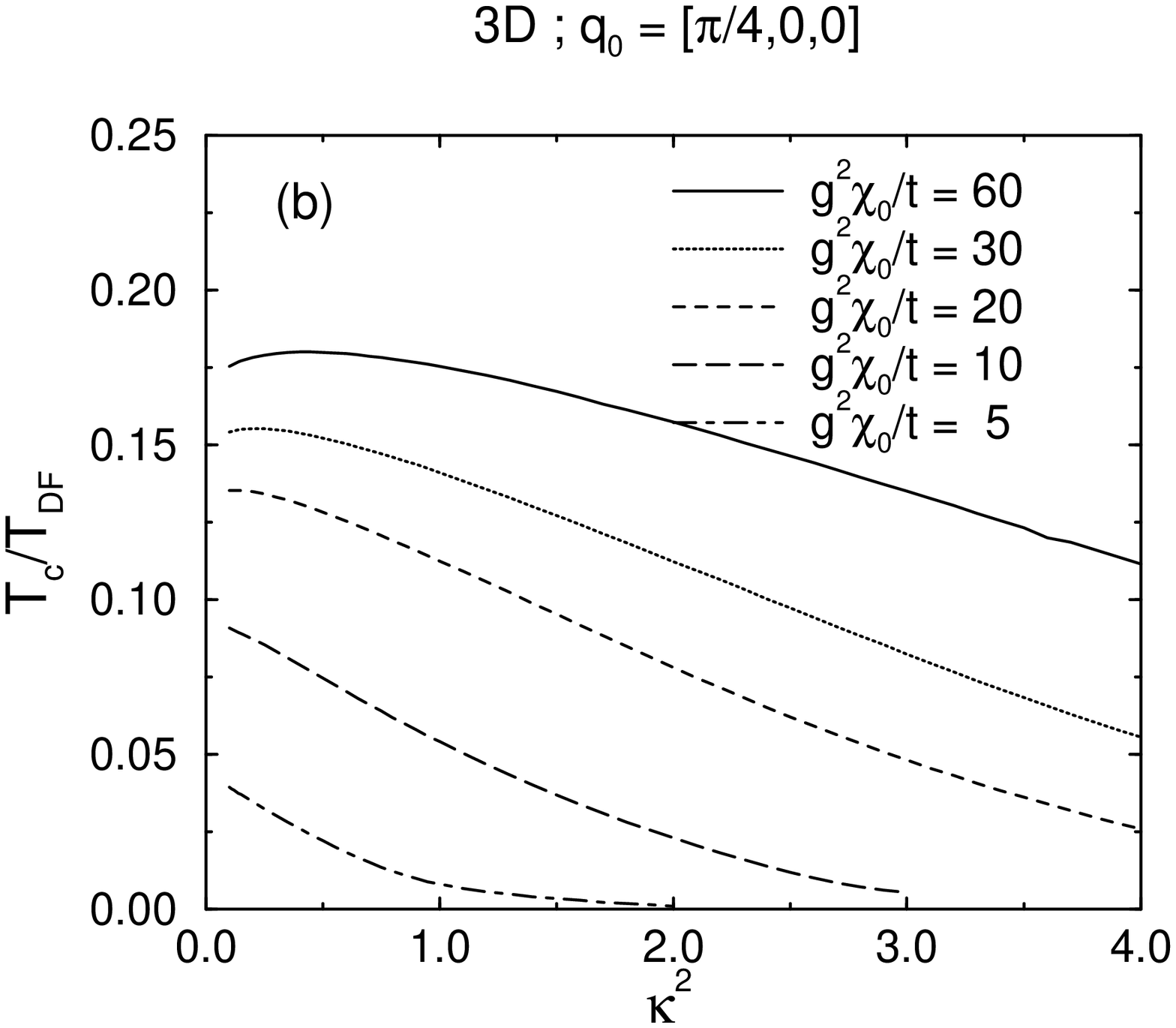}}
\vskip 1.0cm
\caption{(a) and (b) show the spin-singlet $d_{x^2-y^2}$ 
Eliashberg $T_c/T_{DF}$ for a three-dimensional system with incipient 
ordering wavevector ${\bf q_0} = [\pi/4,0,0]$ in as a function of 
the coupling constant $g^2\chi_0/t$ (a) and correlation wavevector 
$\kappa^2$ (b) for the choice $T_{DF} = 2t/3$ and $\kappa_0^2 = 12$. }
\label{fig7}
\end{figure}

\begin{figure}
\centerline{\epsfysize=6.00in
\epsfbox{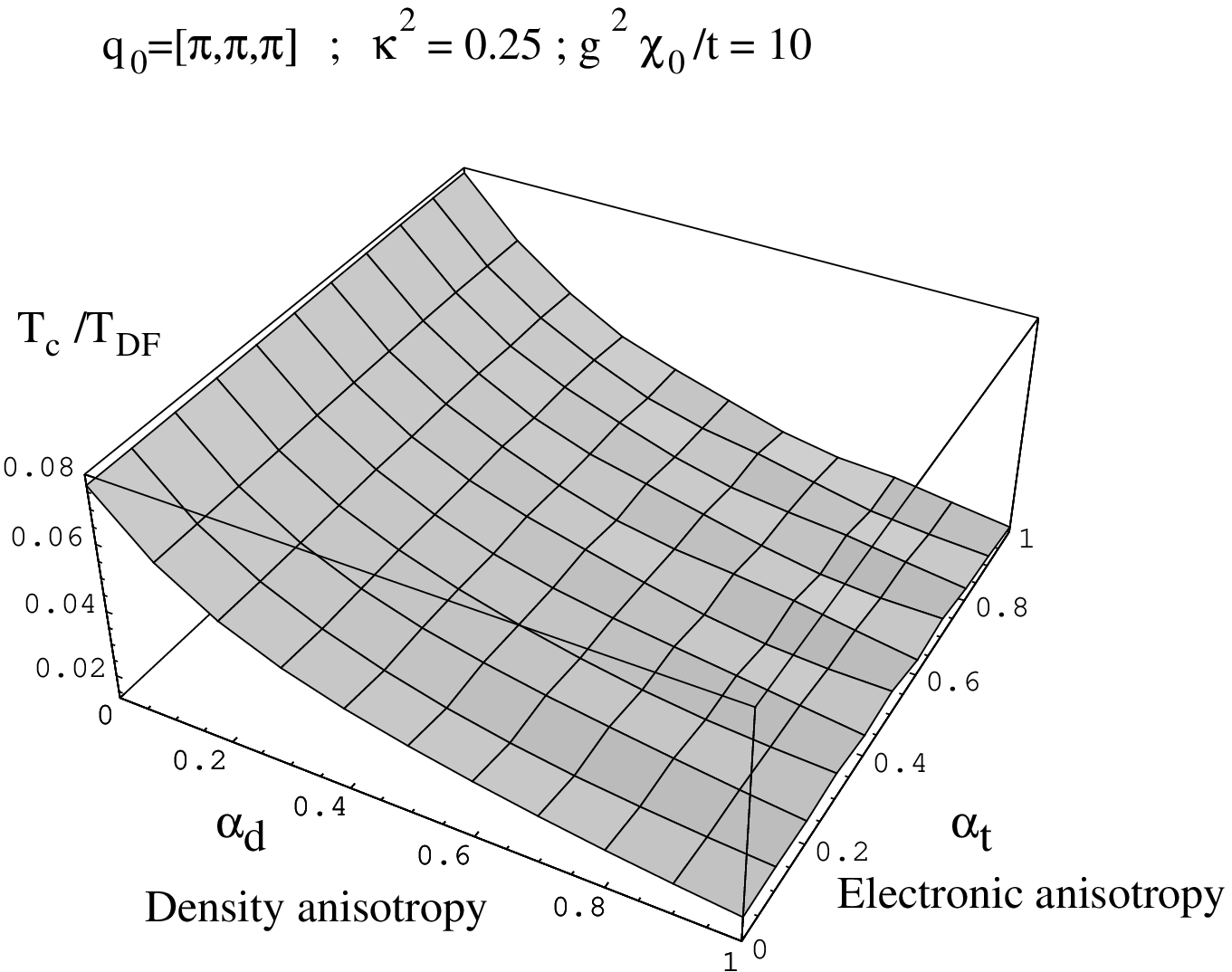}}
\vskip 1.0cm
\caption{Spin-singlet $d_{xy}$ Eliashberg $T_c/T_{DF}$ as a function
of the density and electronic anisotropy parameters $\alpha_d$ and
$\alpha_t$ respectively. $\alpha_d = \alpha_t = 0$ corresponds to the 
2D limit while $\alpha_d = \alpha_t = 1$ corresponds to an isotropic 
3D system. The incipient ordering wavevector is 
${\bf q_0} = [\pi,\pi,\pi]$, and the other model parameters are
$\kappa^2=0.25$, $g^2\chi_0/t = 10$, $T_{DF} = 2t/3$ and 
$\kappa_0^2 = 12$.}
\label{fig8}
\end{figure}

\begin{figure}
\centerline{\epsfysize=6.00in
\epsfbox{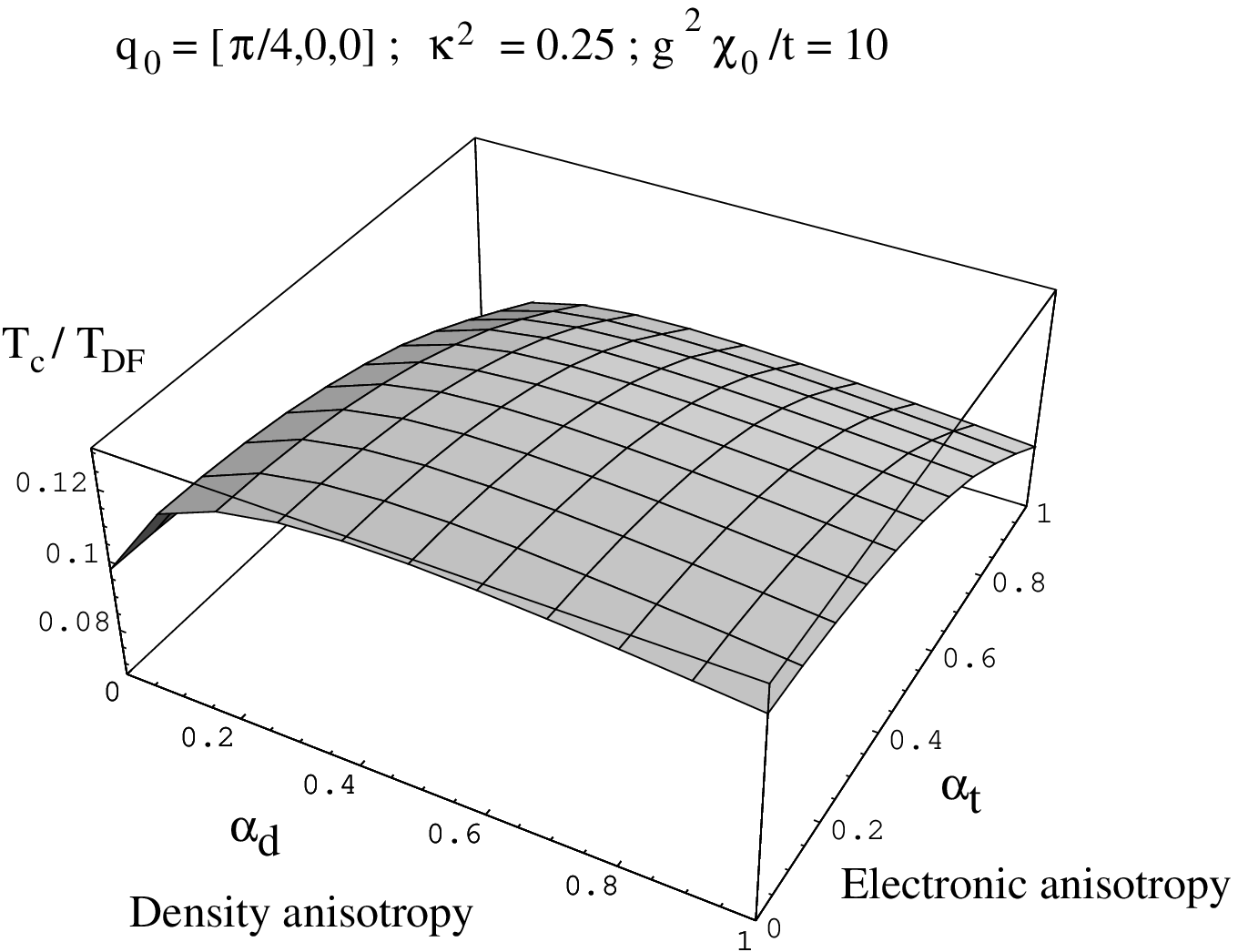}}
\vskip 1.0cm
\caption{Spin-singlet $d_{x^2-y^2}$ Eliashberg $T_c/T_{DF}$ as a function
of the density and electronic anisotropy parameters $\alpha_d$ and
$\alpha_t$ respectively. $\alpha_d = \alpha_t = 0$ corresponds to the 
2D limit while $\alpha_d = \alpha_t = 1$ corresponds to an isotropic 
3D system. The incipient ordering wavevector is 
${\bf q_0} = [\pi/4,0,0]$, and the other model parameters are
$\kappa^2=0.25$, $g^2\chi_0/t = 10$, $T_{DF} = 2t/3$ and 
$\kappa_0^2 = 12$.}
\label{fig9}
\end{figure}

\begin{figure}
\centerline{\epsfysize=6.00in
\epsfbox{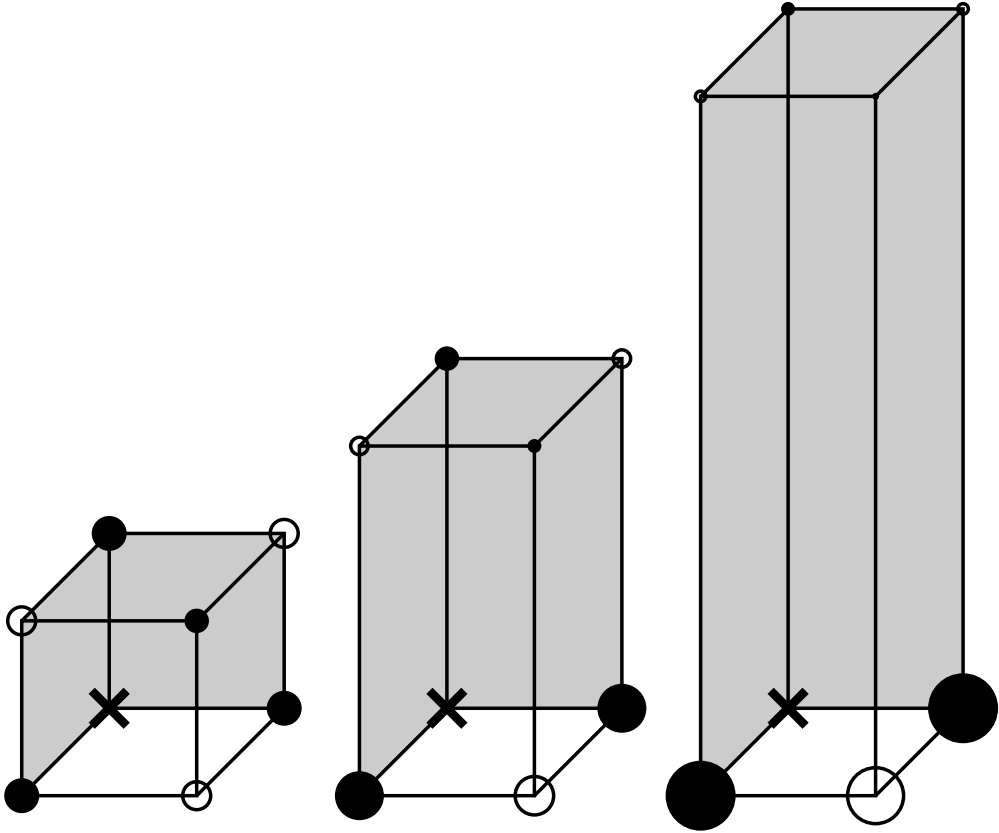}}
\vskip 1.0cm
\caption{The pairing potential for ${\bf q_0} = [\pi,\pi,\pi]$
seen by a quasiparticle in a spin-singlet $d_{xy}$ Cooper pair 
state given that the other quasiparticle is at the origin 
(marked by a cross).  The figure depicts the evolution of 
the potential as one goes from a cubic to a tetragonal 
lattice by varying the parameter $\alpha_d$.  Closed circles 
denote repulsive sites and open circles attractive ones.  
The size of the circle is a measure of the strength of the 
interaction. The nodal planes of the $d_{xy}$ state are 
represented by the shaded region.}
\label{fig10}
\end{figure}

\begin{figure}
\centerline{\epsfysize=6.00in
\epsfbox{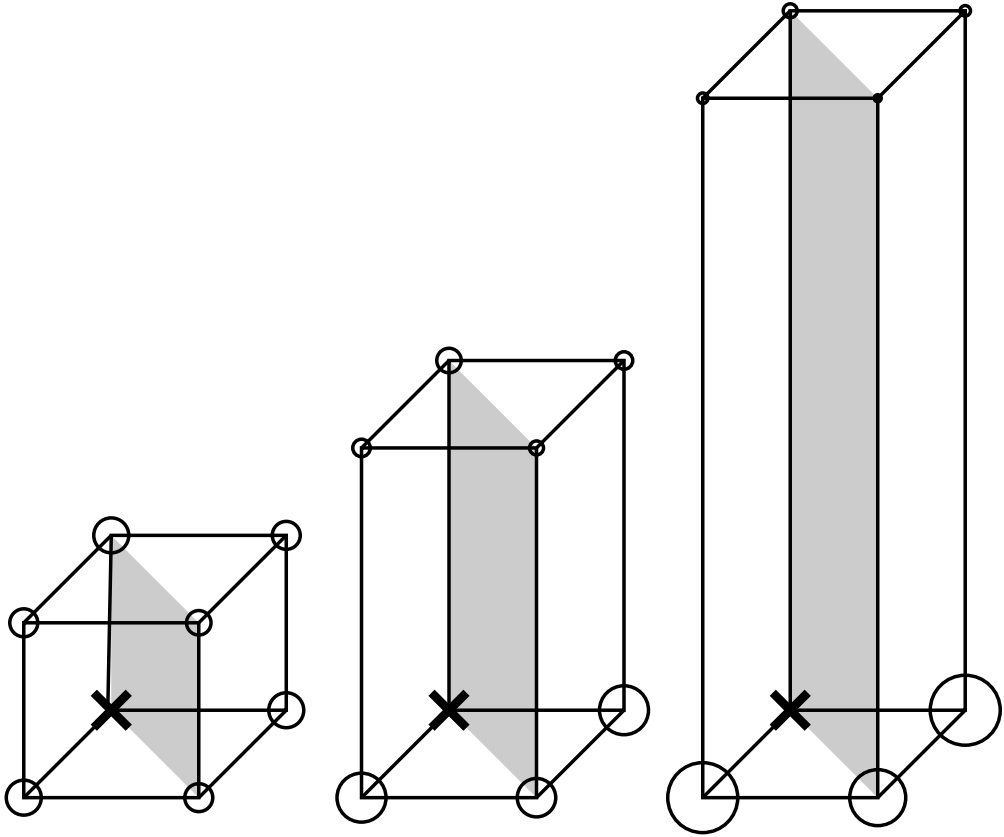}}
\vskip 1.0cm
\caption{The pairing potential for ${\bf q_0} = [\pi/4,0,0]$ 
seen by a quasiparticle in a spin-singlet $d_{x^2-y^2}$ Cooper pair 
state given that the other quasiparticle is at the origin 
(marked by a cross).  The figure depicts the evolution of 
the potential as one goes from a cubic to a tetragonal 
lattice by varying the parameter $\alpha_d$.  Open circles 
denote attractive sites. The size of the circle is a measure 
of the strength of the interaction.  The nodal plane of
the $d_{x^2-y^2}$ state is represented by the shaded region.}
\label{fig11}
\end{figure}

\end{document}